\documentclass[twocolumn,superscriptaddress,showpacs,prd,aps,amsmath,amssymb,nofootinbib]{revtex4-1}                                        
\usepackage{natbib}
\usepackage{graphicx,color}
\usepackage{amsmath,amssymb}
\usepackage{verbatim}
\usepackage{float}
\usepackage{wasysym}
\usepackage{amssymb,graphicx}
\usepackage{epsfig}
\usepackage{psfrag}
\usepackage{dsfont}
\usepackage{amsfonts}
\usepackage{mathrsfs}
\usepackage{multirow}
\usepackage{times}
\usepackage{bm}
\usepackage{hyperref}
\usepackage{scalefnt}
\usepackage{xspace}
\hypersetup{
  colorlinks=true,        
  linkcolor=blue,         
  citecolor=cyan,         
}
\usepackage{pifont}
\usepackage[normalem]{ulem}   
\usepackage{grffile}

\graphicspath{{./}}

\newcommand{\GA}{\alpha}
\newcommand{\GB}{\beta}
\newcommand{\GG}{\gamma}

\newcommand{\GE}{\epsilon}

\newcommand{\GR}{\rho}
\newcommand{\GS}{\sigma}
\newcommand{\GT}{\tau}
\newcommand{\GC}{\psi}
\newcommand{\GX}{\chi}
\newcommand{\GO}{\omega}

\newcommand{\GP}{\phi}

\newcommand{\GU}{\theta}

\newcommand{\ks}{\rm \scriptscriptstyle KS}
\newcommand{\pd}{\partial}
\newcommand{\be}{\begin{equation}}
\newcommand{\ee}{\end{equation}}

\newcommand{\TD}[2]{\tilde{#1}_{#2}}
\newcommand{\TU}[2]{\tilde{#1}^{#2}}
 
\newcommand{\TDD}[3]{\tilde{#1}_{#2 #3}}

\newcommand{\TWD}[2]{( \tilde{\mathbb L} \tilde{W} )_{#1 #2}}

\newcommand{\tA}{\tilde{A}}

\newcommand{\tW}{\tilde{W}}
\newcommand{\tGS}{\tilde{\GS}}
\newcommand{\tGG}{\tilde{\GG}}

\newcommand{\fD}{{\raise0.8ex\hbox{${}^{\ \scalebox{0.7}{$\circ$}}$}}\hspace{-6.6pt}D}

\newcommand{\illinois}{\textsc{\scalefont{1.1}Illinois grmhd}\xspace}

\newcommand{\mbh}{M_{\rm bh}}

\def\QEQ{{%
    \setbox0\hbox{$I$}%
    \rlap{\hbox to \wd0{\hss--\hss}}\box0
}}

\begin{document}

\title{Self-gravitating disks around rapidly spinning, tilted black holes: General relativistic simulations}

\author{Antonios Tsokaros}
\affiliation{Department of Physics, University of Illinois at Urbana-Champaign, Urbana, IL 61801, USA}
\email{tsokaros@illinois.edu}

\author{Milton Ruiz}
\affiliation{Department of Physics, University of Illinois at Urbana-Champaign, Urbana, IL 61801, USA}
\affiliation{Departamento de Astronom\'{\i}a y Astrof\'{\i}sica, Universitat de Val\`encia,
  Dr. Moliner 50, 46100, Burjassot (Val\`encia), Spain}
\author{Stuart L. Shapiro}
\affiliation{Department of Physics, University of Illinois at Urbana-Champaign, Urbana, IL 61801, USA}
\affiliation{Department of Astronomy \& NCSA, University of Illinois at Urbana-Champaign, Urbana, IL 61801, USA}
\author{Vasileios Paschalidis} 
\affiliation{Departments of Astronomy and Physics, University of Arizona, Tucson, AZ 85719}

\date{\today}

\begin{abstract}
We perform general relativistic simulations of self-gravitating black
hole-disks in which the spin of the black hole is significantly
tilted ($45^\circ$ and $90^\circ$) with respect to the angular momentum of the 
disk and the disk-to-black hole mass ratio is $16\%-28\%$. 
The black holes are rapidly spinning with dimensionless spins up 
to $\sim 0.97$. These are the first self-consistent hydrodynamic
simulations of such systems, which can be prime sources for
multimessenger astronomy. In particular tilted black hole-disk systems lead to:
i) black hole precession; ii) disk precession and warping around the black hole;
iii) earlier saturation of the Papaloizou-Pringle instability compared to 
aligned/antialigned systems, although with a shorter mode growth timescale;
iv) acquisition of a small black-hole kick velocity; v) significant
gravitational wave emission via various modes beyond, but as strong as, the typical 
$(2,2)$ mode; and  vi) the possibility of a broad alignment of the angular momentum of
the disk with the black hole spin. This alignment is not related to the
Bardeen-Petterson effect and resembles a solid body rotation.  Our simulations
suggest that any electromagnetic luminosity from our models may power relativistic
jets, such as those characterizing short gamma-ray bursts.
Depending on the black hole-disk system scale the gravitational waves may be
detected by LIGO/Virgo, LISA and/or other laser interferometers.
\end{abstract}

\maketitle

\section{Introduction}
\label{sec:intro}

Black holes (BHs) immersed in gaseous environments are ubiquitous in the
Universe.
Black hole-disks (BHDs) appear on a great variety of scales, reflecting
their diverse birth channels and sites. From the core collapse of massive stars
\cite{Woosley1993,MacFadyen1999} and the cores of active galactic nuclei
\cite{LyndenBell1969,Shakura1973,Paczynski1978}, to asymmetric supernova
explosions in binary systems \cite{Fragos2010}, and the merger of compact
binaries where at least one of the companions is not a BH, BHDs may be 
formed and serve as prime candidates for
multimessenger astronomy.

The magnitude of the spin of the BH, as well as its orientation relative to the
fluid flow, can have large effects, as in the existence and geometry of a
relativistic plasma jet (see e.g. \cite{Liska2018}).  This jet, which can be  powered either by
magnetic fields threading the event horizon and extracting rotational energy
from the BH \cite{BZeffect77}, or from the accretion flow
\cite{1982MNRAS.199..883B}, can precess when misalignment between the BH and
disk angular momentum arises \cite{Aalto2016,Abraham2018,Liska2018}.  Such
misalignment is expected to be a common phenomenon \cite{Fragile2001} both in
active galactic nuclei as well as in BH X-ray binaries
\cite{Hjellming1995,Greene2001,Maccarone2002,Caproni2006,Fragos2010,Aalto2016,Abraham2018,Russell2019}.
Even in the recent observation of M87 by the Event Horizon Telescope
\cite{Akiyama2019_L1} misalignment could not be excluded
\cite{Chatterjee2020,Park2018}.

Tilted BHDs are also the outcome from stellar-mass compact object collisions
when their individual spins are not aligned with the orbital angular momentum
\cite{Foucart:2010eq,Foucart:2012vn,Kawaguchi2015,Dietrich2017c,Chaurasia2020}.
Population synthesis studies suggest that in approximately half of the
BH-neutron star binaries the angle between the orbital angular momentum and the
BH spin is larger than $45^\circ$ \cite{Belczynski08}. Such systems will yield
misaligned BHDs which in turn will affect the existence and the properties of
an electromagnetic counterpart, such as a short gamma-ray burst or a kilonova.

Central to the analysis of a tilted BHD is the so-called Lense-Thirring (LT)
precession \cite{LenseThirring1918}, a gravitomagnetic (GM) effect, according to 
which frame-dragging produced by the rotating and tilted BH causes precession 
of a test ring with angular velocity  
$\Omega_{\rm GM-ring}\approx 2GJ_{\rm bh}/(c^2 r^3)$,
where $J_{\rm bh}$ is the BH angular momentum, and $r$ the ring radius.
In the presence of viscosity (as, for
example, created by a magnetic field) the cumulative effect of LT precession and
internal disk viscosity torques, is the alignment of the angular momenta of the
BH and the disk, a phenomenon known as the Bardeen-Petterson (BP) effect
\cite{Bardeen1975}. Due to the rapid fall-off behavior of the LT angular
velocity, this alignment only affects the inner parts of the disk, within the
so-called Bardeen-Petterson radius, while the outer parts keep their initial
orientation.  The LT and BP effects have been invoked to explain the
quasiperiodic oscillations \cite{vanderklis2005} observed in the X-ray
brightness of a number of neutron star and BH X-ray binaries
\cite{Stella1998,Markovic1998,Fragile2001,Ingram2009}. 
Similarly the GM field will make the BH precess around the disk's rotation
axis. This effect has been invoked
to explain the precession of jets in tidal disruption events (where a star is
tidally disrupted by a supermassive BH) \cite{Stone2012}.  Even in the absence
of a jet, the precession of such disks may have observable consequences.

In general, BHD systems (tilted or not) are subject to various instabilities
that can lead to significant accretion and ablate away the disk. One such
instability is the so-called dynamical runaway instability \cite{Abramowicz83}
where the overflow of a potential surface (similar to the Roche lobe) by the
disk matter will lead to a cascading instability and the final consumption of
the disk by the BH \cite{Font02a,Daigne04,Korobkin2013}. In binary mergers
where a BHD is the final remnant, it was found
\cite{Rezzolla:2010,Hotokezaka2013} that the axisymmetric runaway instability
is of limited importance due to the power-law dependence of the specific
angular momentum profile of the disk \cite{Daigne04}. Therefore its influence
in the formation of ultrarelativistic jets is probably negligible
\cite{prs15,Ruiz:2016rai,Ruiz:2018wah,Ruiz2021}.

A less dramatic instability was discovered by Papaloizou and Pringle
\cite{Papaloizou84} that transports angular momentum outwards and leads to the
formation of an one-arm instability, the so-called Papaloizou-Pringle
instability (PPI).  Using perturbation theory, the authors found a quartic
algebraic equation for the angular velocity of the perturbation mode whose
solutions contain 2 stable modes (real solutions) and 2 unstable ones
(imaginary solutions). These wave perturbations depend on the inner and outer radii of the
disk \cite{Blaes1986,Balbus2003} and highlight the importance of these boundaries
in the development of the PPI. The instability manifests itself when the a wave
which is traveling backwards relative to the fluid at the inner edge exchanges
energy and angular momentum with the wave which is traveling forwards relative
to the fluid at the outer edge.  Angular momentum is transferred outwards, making
the wave at the outer edge that has positive angular momentum grow in
amplitude while the one in the inner edge that has negative angular momentum 
also grow in amplitude, since it is losing angular momentum
\cite{Papaloizou85,Zurek1986,Goldreich1986,Blaes1987,Hawley1987,Goodman1987,
Hawley91,Papaloizou1995,Goodman2001,Heinemann2012}.
A similar mechanism in rotating stars leads to the
Chandrasekhar–Friedmann–Schutz instability
\cite{Chandrasekhar1970a,Friedman1978a,Friedman1978c} which is induced by
gravitational radiation.  The PPI, which was originally found in constant
specific angular momentum disks, can also be developed in BHDs with a nonconstant
specific angular momentum ($\ell$) profile \cite{Papaloizou85}.  Newtonian
analysis finds disks with $\ell\sim r^q$ where $q<2-\sqrt{3}=0.266$  to be
unstable, where the critical exponent $q$ could be even smaller, i.e.  $q\sim 0.25$
\cite{Zurek1986}.  In general the growth of the nonaxisymmetric instability is
more efficient for a smaller exponent $q$ \cite{Zurek1986,Balbus2003}.
Accretion onto the BH has a stabilizing effect on the PPI since the waves at
the inner boundary are disturbed \cite{Blaes1987,Hawley91,DeVilliers2002}. This
is especially true for wide disks, while in more slender ones the PPI seems to
be less affected \cite{Blaes1988}.

The first full general relativistic simulations of a tilted thick disk onto a
Kerr BH \cite{Fragile2005} have demonstrated that LT precession results in a
torque that tends to twist and warp the disk, similar to Newtonian studies
\cite{Nelson2000}.  The authors found that this precession depends primarily on
the sound speed in the disk.  For disks where in their bulk the LT timescale
was less than the azimuthal sound crossing time, the disk undergoes
differential precession out to a transition radius. On the other hand when the
the LT timescale was greater than the azimuthal sound crossing time, the disk
undergoes near rigid-body precession after a short initial period of
differential precession.  Another interesting finding in \cite{Fragile2005} was
the tendency for these disks to align toward the equatorial plane of the BH,
despite the lack of viscous angular momentum transport. According to the
authors this alignment between the angular momentum of the disk and the BH spin
was facilitated by the preferential accretion of highly tilted disk material
that resulted in the depletion of the misaligned disk angular momentum. Since
the authors considered disks with mass much smaller than the BH (test-fluid
limit) the spin of the BH was unaffected.  Such kind of purely hydrodynamical
alignment has also been found in BH-neutron star simulations
\cite{Kawaguchi2015}, where the alignment timescale was of the same order as
the disk precession timescale.  The authors speculated that 
this BP-like behavior is induced by a purely hydrodynamical mechanism, such
as angular momentum redistribution due to a nonaxisymmetric shock wave excited 
in the disk\footnote{
Notice that in the numerical simulations of \cite{Nealon2015} using a 
post-Newtonian description of the central potential and an artificial 
viscosity, the BP picture of an aligned inner disk occurred only at low 
inclinations and only when Einstein precession was not accounted for. In high 
resolution calculations with the Einstein precession included, the authors 
found steady-state oscillations in the disk tilt, as well as the breaking of the
disks that are relatively thin and highly misaligned to the BH spin 
\cite{Ivanov1997,Demianski1997,Ogilvie1999,Nelson2000,Lubow2002}.}.

The assumption that  the mass of the disk is negligible in comparison with the
mass of the central BH may not always be valid.  Some isolated or binary BHs
detectable by LISA may find themselves immersed in extended disks with masses
comparable or greater than the BHs themselves. This may be particularly true of
stellar-mass BHs in AGNs and quasars or supermassive BHs in extended
disks formed in nascent or merging galactic nuclei. The gravitational pull of
the disk on the binary can be important in such cases, the accretion rate from
the inner disk radius can be high and even super-Eddington, orbital and spin
precession as well as spin flipping in the case of misaligned disks is a
possibility, while density perturbations in the disk can arise from
instabilities. Alternative scenarios for the formation of massive BHDs include
the collapse of rapidly rotating, supermassive stars or the merger of binary
stellar systems (such as a neutron star-white dwarf) with significant asymmetry
in their mass or spin.  In binaries the mass of the disk depends on how far from
the BH is the secondary compact object being disrupted \cite{Foucart:2012nc}.
If tidal disruption happens far from the innermost stable circular orbit (ISCO)
of the BH, then a disk with a large mass is produced. On the other hand, small
mass disks (or even essentially no disk at all) are produced when tidal
disruption happens close to the ISCO of the BH (or inside it).  This crucial
distance that controls the importance of self-gravitation for the disk 
depends on the mass
ratio of the binary, the compactness of the primary and the BH spin. The mass
of the disk increases with a larger BH spin (since the ISCO decreases with
increasing spin) and decreases with a larger BH mass (the ISCO increases with
increasing BH mass) \cite{Rezzolla:2010,Lovelace:2013vma}.

Only by including self-gravity in full general relativity and tracking the
nonaxisymmetric perturbations that self-gravity may trigger can gravitational
waves from the disk be calculated reliably. Such perturbations and
gravitational waves can be detected by LISA and other instruments
\cite{Montero2010,Kiuchi:2011re,Mewes2015,Mewes2016,Wessel:2020hvu,Shibata:2021sau}.
Also, disk self-gravity must be incorporated to determine the astrophysical
consequences of BH precession, which may, for example, trigger X-shape radio 
galaxies \cite{Ekers1978,Cheung2007,Bera2020}.

General relativistic studies of self-gravitating BHDs have been performed in a
number of works
\cite{Montero2010,Kiuchi:2011re,Korobkin2011,Korobkin2013,Mewes2015,Mewes2016,
Wessel:2020hvu,Shibata2021} and the roles of the runaway instability, as well as
the PPI, have been elucidated. Although most of the
BHDs will not develop the runaway instability (e.g.
\cite{Montero2010,Rezzolla:2010,Kiuchi:2011re,Korobkin2011}), it cannot be
excluded when more favorable circumstances are present \cite{Korobkin2013}
(e.g. disks that fill their Roche lobes). Regarding the PPI, it was found
that, as in Newtonian gravity, self-gravitating BHDs are subject to an $m=1$
nonaxisymmetric mode growth under a wide range of conditions \footnote{ Note
that early studies in Newtonian gravity
\cite{Goodman1988,Papaloizou89,Tohline1990,Christodoulou1992,Christodoulou1993}
have shown that self-gravity inhibits the PPI for all angular momentum
profiles, while new kinds of nonaxisymmetric instabilities arise. These
include the I-mode
(``intermediate'') that leads to fission, and the J-mode (Jeans
instability) that leads to fragmentation. }.  In \cite{Korobkin2011} it was
shown explicitly that the $m=1$ PPI mode is accompanied by an outspiraling
motion of the BH, which further amplifies the one-arm instability.  More
massive tori and a constant specific angular momentum profile favors the
appearance of the PPI, in contrast with less massive disks and/or a
non-constant $\ell$ profile, for which the disk may even be PP-stable
\cite{Kiuchi:2011re}.  In addition since the nonaxisymmetric structure survives
long after the saturation of the PPI, these systems can be promising
sources for coincident detections of electromagnetic and gravitational waves similar
to GW170817.  The above works focused on tori around nonspinning BHs and 
were later extended to BHDs around spinning BHs in
\cite{Mewes2015,Wessel:2020hvu,Shibata2021}.  In \cite{Wessel:2020hvu} it was
speculated that the accretion rate in PPI unstable disks may be used to measure
the BH spin. It was found that systems of $\sim 10 M_\odot$ --relevant for for
BH–neutron star mergers-- will be  detectable by the Cosmic Explorer out to $\sim
300$ Mpc, while DECIGO (LISA) will be able to detect systems of $\sim 1000
M_\odot\ (10^5 M_\odot)$. The latter are relevant for disks forming in collapsing,
supermassive stars out to cosmological redshift of $z\sim 5\ (z\sim 1)$.  In
\cite{Shibata2021} an alternative scenario for event GW190521 was put forward.
In particular it was conjectured that GW190521 may not represent the merger of
binary BHs, but instead the stellar collapse of a very massive star, leading
temporarily to a BH of mass $\sim 50 M_\odot$ and a massive disk of several
tens of solar masses that is dynamically unstable to the PPI.

The first general relativistic simulations where the spin of the BH is tilted
with respect to the angular momentum of the disk were performed in
\cite{Mewes2015,Mewes2016}, albeit starting from artificial initial values. In
particular the authors first computed models of self-gravitating, massive tori
around nonrotating BHs \cite{Stergioulas2011}, and then replaced the resulting
spacetime with a tilted Kerr metric in quasi-isotropic coordinates, while
retaining the hydrodynamical profile. Notwithstanding these initial conditions
the authors performed a thorough investigation of the twist (precession) and
the tilt (inclination) of the disk, finding that for BHD mass ratios of
$\gtrsim 4\%$ the assumption of using a fixed background spacetime is
unjustified. The authors observed significant precession and nutation of the
tilted BH as a result of the disk evolution, which cannot be accounted in fixed
spacetime simulations.  The LT torque that the BH exerts on the disk forces the
disk to precess as a solid body which in turn leads to BH precession.  The
simulations of \cite{Mewes2015,Mewes2016} showed the universal character of the
PPI with regards to initial spin magnitudes, tilt angles, and disk angular 
momentum profiles.

In this work we extend previous studies of self-gravitating BHDs in two ways.
For the first time we perform general relativistic simulations of tilted BHDs
starting from \textit{self-consistent} initial values. The tilted BHD models are
solutions of the full (i.e. including the conformal metric) general
relativistic initial value problem as described in \cite{Tsokaros2018a}.
Second, we extend the parameter space by evolving disks around \textit{rapidly
spinning} BHs (aligned, antialigned and tilted with respect to the disk angular
momentum) having dimensionless spins up to $0.97$. We find that although the
saturation of the PPI appears significantly earlier for tilted BHDs than those
with aligned/antialigned spins, due to the inherent initial nonaxisymmetry,
their growth rate is smaller. The maximum density in the disk can increase by
orders of magnitude, while the disk precesses and warps around the BH. The BH
itself also precesses and its spin can increase or decrease depending the
initial configuration. In one case where the initial BH spin was tilted at
$45^\circ$ with respect to the angular momentum of the disk the BH was spun up
to a maximal value, beyond which we couldn't continue our simulation. In another
case where the initial BH spin was tilted by $90^\circ$ accretion spun down the
BH. By computing the precession timescales we confirmed their agreement with
post-Newtonian estimates. The precessing BHDs are responsible for copious
gravitational wave emission in multiple modes, which we compute. In general the
gravitational wave strain appears to be an order of magnitude larger than
previous calculations
\cite{Kiuchi:2011re,Mewes2015,Wessel:2020hvu,Shibata2021} with a diverse
spectrum. Although our simulations do not include magnetic fields, 
estimation of the effective turbulent magnetic viscous timescale shows that it is 
much longer than the
dynamical timescale of the one-arm instability. 
Therefore we expect these BHDs to be
prominent sources of gravitational waves and 
Poynting electromagnetic radiation (in the presence of magnetic
fields) and thus excellent sources for multimessenger astronomy.

In this paper, spacetime indices are Greek, spatial indices Latin, and we employ 
geometric units in which $G=c=M_\odot=1$, unless stated otherwise.


\section{Initial data}
\label{sec:id}

\setlength{\tabcolsep}{5pt}                                                      
\begin{center}                                                                  
\begin{table*}
\caption{The initial BHD models. The angular momentum of the disks is along the
z-axis.  Columns are the model name, the magnitude of the dimensionless BH spin
$\chi=J_{\rm bh}/M_{\rm bh}^2$, the angles of the spin angular momentum in
spherical coordinates $(\GU_s,\GP_s)$, the inner specific angular momentum $\ell_{\rm in}$, the inner edge of
the disk $r_{\rm in}$, the maximum density coordinate $r_c$, the outer edge of the disk $r_{\rm out}$, the rest
mass of the disk $M_0$, the
Arnowitt-Desser-Misner (ADM) mass $M$, the period of the maximum density point of
the disk $P_c$, the dynamical time $t_d\sim 1/\sqrt{\GR_{\rm max}}$, and the
precession angular velocity $P_{\rm GM}$ of the BH as calculated in Sec. \ref{sec:pf}. 
Here $\mbh$ is the mass of the BH.
Center dots denote ``not applicable''.}                       
\label{tab:id}                                                     
\begin{tabular}{cccccccccccccc}                                                    
\hline\hline                                                              
Model & $\chi$   & $(\GU_s,\GP_s)$       & $\ell_{\rm in}/\mbh$ &  $r_{\rm in}/\mbh$  &  $r_c/\mbh$ & $r_{\rm out}/\mbh$ & $M_0/\mbh$  & $M/\mbh$ & $P_c/\mbh$ 
& $t_d/\mbh$  & $P_{\rm GM}/P_c$  \\ \hline\hline
A1    & $0.966$  & $(0^\circ,0^\circ)$   & $4.63$            &   $10$   &   $17.3$ &   $49.4$    & $0.259$ &  $1.273$ & $462$ & $297$ & $\cdots$ \\ \hline
A2    & $0.957$  & $(45^\circ,0^\circ)$  & $4.60$            &   $10$   &   $17.3$ &   $49.4$    & $0.156$ &  $1.167$ & $435$ & $371$ & $60$ \\ \hline
A3    & $0.968$  & $(90^\circ,0^\circ)$  & $4.85$            &   $10$   &   $17.9$ &   $51.1$    & $0.280$ &  $1.298$ & $455$ & $290$ & $35$ \\ \hline
A4    & $0.963$  & $(180^\circ,0^\circ)$ & $5.13$            &   $10$   &   $20.0$ &   $57.3$    & $0.242$ &  $1.256$ & $520$ & $364$ & $\cdots$ \\ \hline
\end{tabular}                                                               
\end{table*}                                                                   
\end{center}

The initial models of the BHDs considered in this work, models A1-A4 in Table
\ref{tab:id}, have been constructed using the \texttt{COCAL} code and the
method described in \cite{Tsokaros2018a}.  In particular we solve the full
initial value Einstein equations by assuming that the conformal 3-dim metric is
decomposed as $\TDD{\GG}{i}{j}:=f_{ij}+h_{ij}$, where $f_{ij}$ is the flat
metric and $h_{ij}$ the nonflat contributions. The metric on the 3-geometry
$\GG_{ij}$ is related to the conformal metric through
$\GG_{ij}=\GC^4\TDD{\GG}{i}{j}$. The nonflat contributions $h_{ij}$ are
computed alongside the lapse $\GA$, shift $\GB^i$, and the conformal factor
$\GC$, assuming ${\rm det}(\TDD{\GG}{i}{j})={\rm det}(f_{ij})$. One of the new
characteristics of this method is the decomposition of the conformal tracefree
part of the extrinsic curvature as
\be
\TDD{A}{i}{j} = \tilde{A}^{\ks}_{ij} + \tGS\TWD{i}{j}  ,  
\label{eq:ctt}
\ee
where $\tilde{A}^{\ks}_{ij}$ is the conformal Kerr-Schild tracefree part,
$\tW_i$ an unknown spatial vector, $\tGS$ a scalar, and $\tilde{\mathbb L}$ the
conformal Killing operator:
$\TWD{i}{j}=\TD{D}{i}\TD{W}{j}+\TD{D}{j}\TD{W}{i}-\frac{2}{3}\TDD{\GG}{i}{j}\TD{D}{k}\TU{W}{k}$.
Here $\TD{D}{i}$ is the covariant derivative with respect to the conformal
metric $\TDD{\GG}{i}{j}$.  It is assumed that $A_{ij}=\GC^4\tA_{ij}$ and
$\tGS=1/(2\GA)$.  As explained in \cite{Tsokaros2018a}, Eq. (\ref{eq:ctt}) with
the appropriate boundary conditions for $\tW_i$ yields a convergent solution
for the potentials $h_{ij}$, which in addition, can be horizon penetrating.
The price paid for this additional decomposition of the extrinsic curvature is
an extra 3 elliptic equations for the potentials $\tW_i$. For the slicing we
assume Kerr-Schild coordinates with $K=K_{\ks}$ under the gauge $\fD_i
h^{ij}=\fD_i h^{ij}_{\ks}$, with $h^{ij}_{\ks}$ being the exact Kerr-Schild
potentials, and $\fD_i$ the covariant derivative with respect to the flat
metric $f_{ab}$. We set $\pd_t\tGG_{ij}=\pd_t\tA_{ij}=\pd_t K=0$.

\begin{figure}
\begin{center}
\includegraphics[width=0.99\columnwidth]{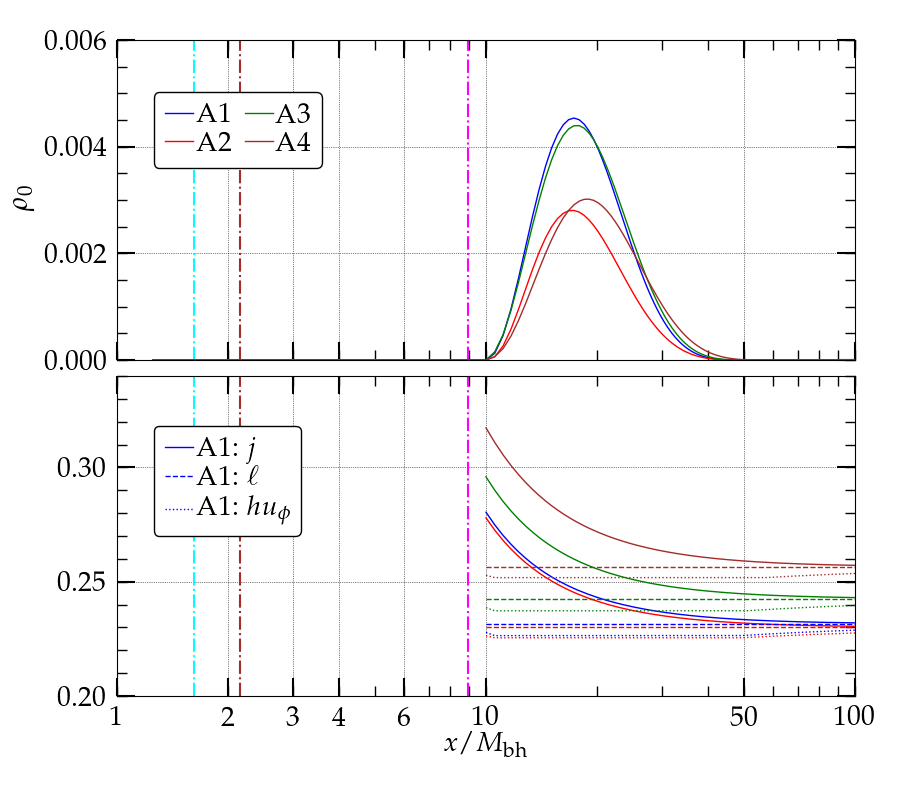}
\caption{Top panel: Initial rest-mass density distribution for the four models
evolved.  Bottom panel: The specific angular momentum of the BHD model A1.
Solid lines correspond to $j$, dashed lines to $\ell$ and dotted lines to
$hu_\GP$.  The vertical dashed dotted lines correspond to the event horizon
(cyan), the marginally stable radius for the prograde orbit (brown), and the
marginally stable radius for the retrograde orbit (magenta) around a BH whose
dimensionless spin is $\GX=0.95$.  }
\label{fig:rhoj}
\end{center}
\end{figure}
\begin{figure}
\begin{center}
\includegraphics[width=0.99\columnwidth]{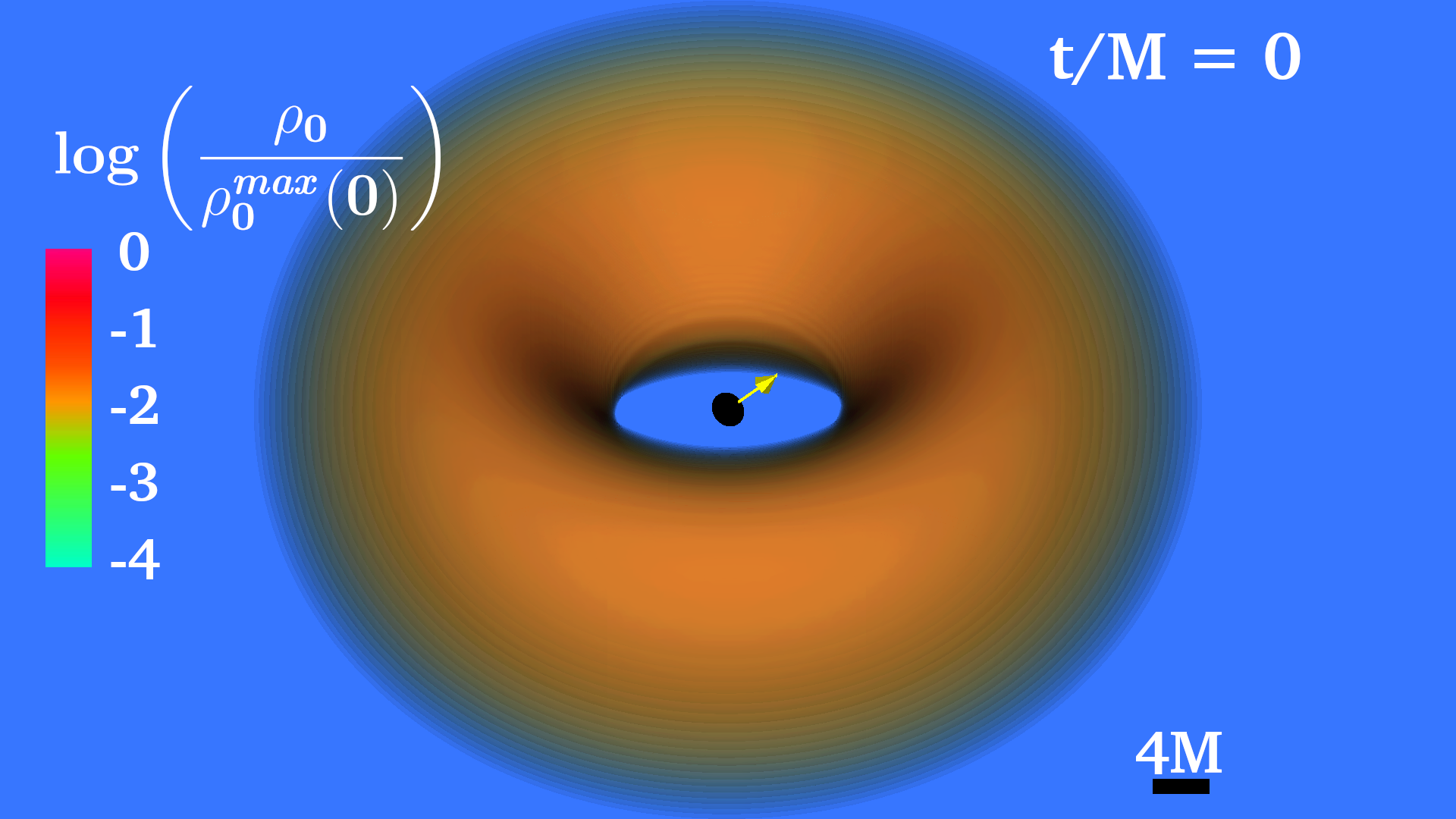}
\caption{Three dimensional rendering of BHD model A2 at $t=0$. The direction of the BH spin 
tilted at $45^\circ$ with respect to the z axis (axis of the orbital angular momentum of the 
disk) is shown by the yellow arrow. The black spheroidal region denotes the apparent horizon.}
\label{fig:t0nocut}
\end{center}
\end{figure}

For the Euler equations we assume stationarity and axisymmetry
\cite{Tsokaros2018a}, which is a reasonable assumption whenever the disk is
far away from the tilted BH.  The density profiles along the x axis for our
models are plotted in the top panel of Fig. (\ref{fig:rhoj}).  The disk is
described by a $\Gamma=4/3$ polytropic equation of state\footnote{This choice
is appropriate for a thermal
radiation-dominated gas, which might be found around a supermassive BH, 
but is not the optimal choice for BH-neutron star binaries.}, having constant
specific angular momentum $\ell=-u_\phi/u_t$.  Note that there exist other
diagnostics for the specific angular momentum, such as $j=u_t u^\GP =
\ell/(1-\Omega \ell)$, as well as $hu_\GP$. Here $h$ is the specific enthalpy,
$u_\GP$ the azimouthal component of the 4-velocity, and $\Omega=u^\GP/u^t$ the
angular velocity of the fluid.  The three diagnostics are plotted in the bottom
panel of Fig. (\ref{fig:rhoj}) for case A1 while similar behavior can be found
for cases A2-A4. Our disk models have both $\ell$ and $hu_\GP$
constant.

For the numerical solution of the Poisson-type of equations we use the
Komatsu-Eriguchi-Hachisu method for BHs, which was first developed in
\cite{Tsokaros2007}.  The self-gravitating BHD is calculated as follows: i)
First we calculate a massless disk \cite{Chakrabarti1985,DeVilliers03b} around
a tilted, spinning BH whose mass is $m$ and dimensionless spin is $a/m=0.95$. We
call $m$ the BH bare mass.  ii) Using as initial data the solution obtained in
(i) we iterate over the Einstein and Euler equations to compute a
self-gravitating disk of a given maximum rest-mass density. iii) By increasing
the maximum density of the disk and repeating step (ii) we compute a sequence
of BHDs whose disk mass is growing. For each solution the angular momentum of
the BH $J_{\rm bh}$ is calculated through the isolated horizon formalism
\cite{IH_DH,Dreyer-etal-2002-isolated-horizons}.  Using the apparent horizon finder
described in \cite{Tsokaros2007} we calculate the mass of the BH $M_{\rm bh} $
\cite{Christodoulou70}, and its dimensionless spin  $\chi=J_{\rm bh}/M_{\rm
bh}^2$. In Fig. \ref{fig:t0nocut} a full three-dimensional rendering of the BHD
model A2 is shown. The yellow arrow depicts the spin of the BH, which is tilted 
at $45^\circ$ with respect to the z axis. The latter coincides with the axis of
rotation of the disk. The apparent horizon is denoted by a black spheroid 
which is similarly tilted. Models A1, A3, and A4 have similar disk structure, 
differing mainly on the tilt angle of the BH.

\subsection{Precession frequencies}
\label{sec:pf}

\begin{figure}
\begin{center}
\includegraphics[width=0.99\columnwidth]{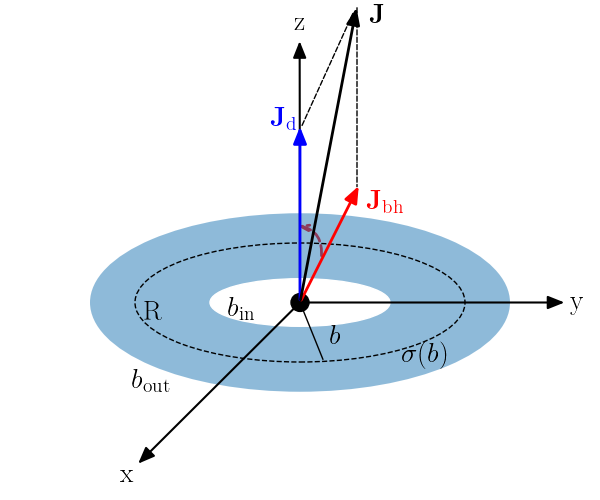}
\caption{A thin massive disk on the xy plane with angular momentum
$\mathbf{J}_{\rm d}$ along the z axis rotates around a tilted spinning BH with
angular momentum $\mathbf{J}_{\rm bh}$. Both of them undergo GM 
precession about the total angular momentum $\mathbf{J}$. }
\label{fig:disk}
\end{center}
\end{figure}

The relevant post-Newtonian (PN) 
theory for understanding a massive disk around a tilted BH is
summarized in \cite{Thorne86}, which we closely follow in the analysis below. In
particular we assume a massive thin disk confined on the xy plane having
angular momentum $\mathbf{J}_{\rm d}$ along the z-axis, and whose inner radius
is $b_{\rm in}$ while its outer radius is $b_{\rm out}$ (see Fig.
\ref{fig:disk}). The disk rotates about a BH having angular momentum
$\mathbf{J}_{\rm bh}$ tilted with respect to $\mathbf{J}_{\rm d}$.  We further
assume that the disk lies outside the BP radius so that it is not driven down
to the hole's equatorial plane (perpendicular to $\mathbf{J}_{\rm bh}$). In our
simulations there is no viscosity, so in principle there is no such accretion
and no BP effect\footnote{ As we discussed in the Introduction, in
\cite{Kawaguchi2015} the authors found such alignment in pure hydrodynamical
simulations. In any case, even if numerical viscosity is present we assume that the bulk
of the mass and angular momentum of the ambient disk remains largely intact (apart
from precession).}.  Now we imagine that the thin disk is composed of massive
rings, each one of them having a mass $dM_{\rm R}$, and radius $b$.  The ring's
GM field  will make the BH precess around the disk's rotation
axis $d\mathbf{J}_{\rm bh}/dt = \boldsymbol{\Omega}_{\rm
GM}\times\mathbf{J}_{\rm bh}$ where $\Omega_{\rm GM} = 2J_{\rm R}/b^3$  and
$J_{\rm R}$ the angular momentum of the ring.  Generalizing to the disk of Fig.
\ref{fig:disk} we can write
\be
d \Omega_{\rm GM} = \frac{2}{b^3} dJ_{\rm R}(b).
\label{eq:dome}     
\ee
If $\sigma(b)$ is the surface gas density and $\omega(b)$ the angular velocity of the ring,
we have
\be
dJ_{\rm R}(b) = \omega(b) b^2 dM_{\rm R} = \omega(b) b^2 (\sigma(b)\; 2 \pi b db),
\label{eq:dj}
\ee
where
\be
\sigma(b) =  \int_{-h(b)}^{h(b)} \rho_0(b,z) dz,\quad
\omega(b) =  \int_{-h(b)}^{h(b)} \Omega(b,z) dz,
\label{eq:sigma}
\ee
are calculated as quadratures over the disk height $h(b)$ at the particular
radius $b$.  In Eq. (\ref{eq:sigma}) $\rho_0(b,z)$ is the rest-mass density of our
3d disks, and $\Omega(b,z)$ their angular velocity profile.  Note that
although in Newtonian gravity von Zeipel's theorem \cite{vonZeipel1924} states
that for a barotropic fluid the angular velocity of a stationary disk 
depends only on the distance
from the axis of rotation (Poincar\'e-Wavre
\cite{Tassoul-1978:theory-of-rotating-stars}), in general relativity the
surfaces of constant $\Omega$ have cylindrical topology, therefore they depend
not only on the distance from the rotation axis but also on the distance from
the equatorial plane \cite{Abramowicz1974,Karkowski2018}.

From Eqs. (\ref{eq:dome})-(\ref{eq:sigma}) the GM precession angular velocity
of the BH will be
\be
\Omega_{\rm GM} =  \int_{b_{\rm in}}^{b_{\rm out}}  4 \pi \omega(b) \sigma(b) db ,
\label{eq:omeGM}
\ee
where $b_{\rm in}$ and $b_{\rm out}$ are the radial boundaries of the disk.
Inserting in Eqs. (\ref{eq:sigma}), (\ref{eq:omeGM}) the density and angular
velocity of our tilted self-gravitating disk models A2 and A3 we can compute
$\Omega_{\rm GM}$. These theoretical PN estimates 
are reported in the last column of Table \ref{tab:id}
in terms of the GM precession period $P_{\rm GM}=2\pi/\Omega_{\rm GM}$.

Note that a ring of mass $M_{\rm R}$ rotating with Keplerian angular velocity
around a BH of mass $M_{\rm bh}$ at a radius $b_{\rm R}$ will be subject to
GM precession with 
\be
M \Omega_{\rm GM} = 2 \left(\frac{M}{b_{\rm R}}\right)^{5/2} 
                      \left(\frac{M_{\rm bh}}{M}\right)^{1/2} 
                      \left(\frac{M_{\rm R}}{M}\right) ,
\label{eq:omeGMkep}
\ee 
where $M$ is the ADM mass of the system. For our models $\{$A2, A3$\}$ Eq.
(\ref{eq:omeGMkep}) yields $P_{\rm GM}/P_c =\{ 54, 31 \}$ in rough agreement
with the values shown in Table \ref{tab:id}.  This shows that despite the
constant specific angular momentum our self-gravitating disks are effectively
close to the Keplerian test-ring model.

Not only does the disk makes the BH to precess: conservation of the total angular
momentum $\mathbf{J}=\mathbf{J}_{\rm bh}+\mathbf{J}_{\rm d}$ implies that the
BH will make the disk  precess, i.e.,
\be
\frac{d\mathbf{J}_{\rm d}}{dt} = \left(\frac{2\mathbf{J}_{\rm bh}}{b^3}\right) \times \mathbf{J}_{\rm d}.
\label{eq:djdt}
\ee
The precession frequency of the disk $\Omega_{\rm GM-disk}$ is related to the
precession frequency of the BH $\Omega_{\rm GM}$ as
\be
\Omega_{\rm GM-disk} = \Omega_{\rm GM} \frac{J_{\rm bh}}{J_{\rm d}} .
\label{eq:omeGMdisk}
\ee
For models A2 and A3 we find that $P_{\rm GM}/P_{\rm GM-disk}$ is of order
$1.0$ implying that the spin of the BH will precess at the same timescale as
the warping of the disk.

\begin{table*}
\caption{Grid parameters used for the evolution of the BHDs of Table
\ref{tab:id}. The computational grid consists of a set of 13 nested refinement
boxes centered on the BH apparent horizon.  The step interval in the coarser
level is $\Delta x_{\rm max}=50\mbh$, while in the finer refinement level is
$\Delta x_{\rm min} \approx 0.0122\mbh$. Note that the ADM mass $M\approx 1.2\mbh -1.3\mbh$
depending on the model. 
}      
\label{tab:evol}                                                               
\scalebox{0.95}{
\begin{tabular}{ccccccccc}
\hline
\hline
$\{x,y,z\}_{\rm min}$ & $\{x,y,z\}_{\rm max}$ &
Grid hierarchy (Box half-length) \\  
\hline                                                              
$-4000\mbh$ & $4000\mbh$ & $\{ 0.5,\; 1.56,\; 3.12,\; 6.24,\; 12.48,\; 25,\; 50,\; 100,\; 200,\; 399,\; 799,\; 1597,\; 4000 \}\mbh$ \\
\hline                                                            
\hline                                                             
\end{tabular}                                                                    
}
\end{table*}

As a final note we mention that the disk's tidal field will also exert a torque
on the BH that leads to tidally torqued precession with angular velocity
\cite{Thorne86}
\be
\Omega_{\rm T} = \frac{3aM_{\rm R}}{2b^3_{\rm R}} \cos\GU_s  .
\label{eq:omeT}
\ee
For model A2 we find $P_{\rm T}/P_{\rm GM}\approx 8$ using $b_{\rm R}$ as the
radius of the maximum density. On the other hand we can perform an analysis
similar to the GM frequency and write 
$d\Omega_{\rm T}=3 a dM_{\rm R}\cos\GU_s /(2b^3)$, with $dM_{\rm R}=\GS(b)2\pi b db$.
Integrating as in Eq. (\ref{eq:omeGM}), we find $P_{\rm T}/P_{\rm GM}\approx 9$  
in agreement with the cruder estimate above. Therefore the tidally
torqued precession is secondary to the GM precession and needs very long
evolutions to be probed.

\begin{figure*}
\begin{center}
\rotatebox{90}{\bf\Large $\phantom{mmm}$Tilt: $0^\circ$ }
\includegraphics[width=0.95\columnwidth]{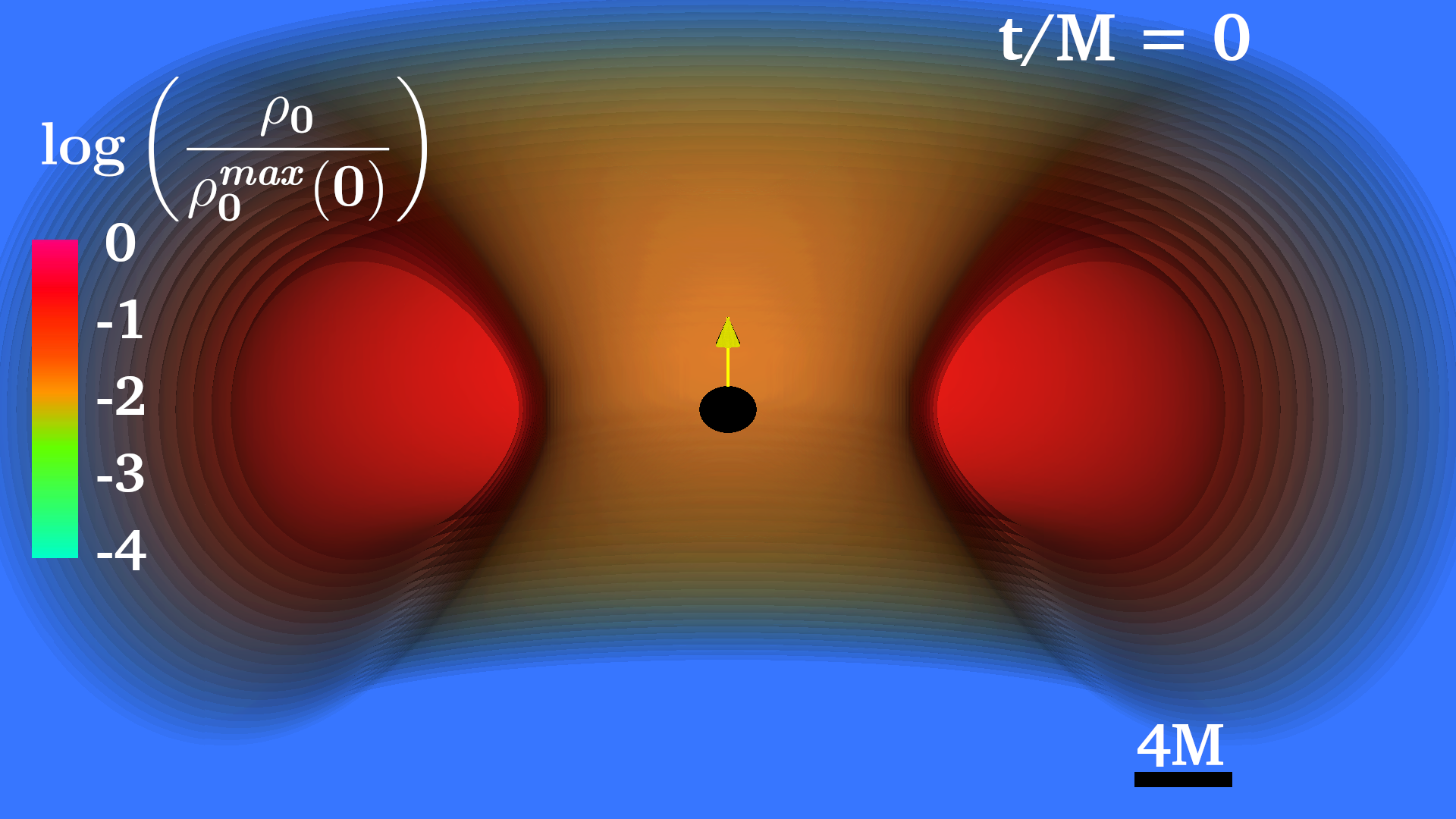}
\includegraphics[width=0.95\columnwidth]{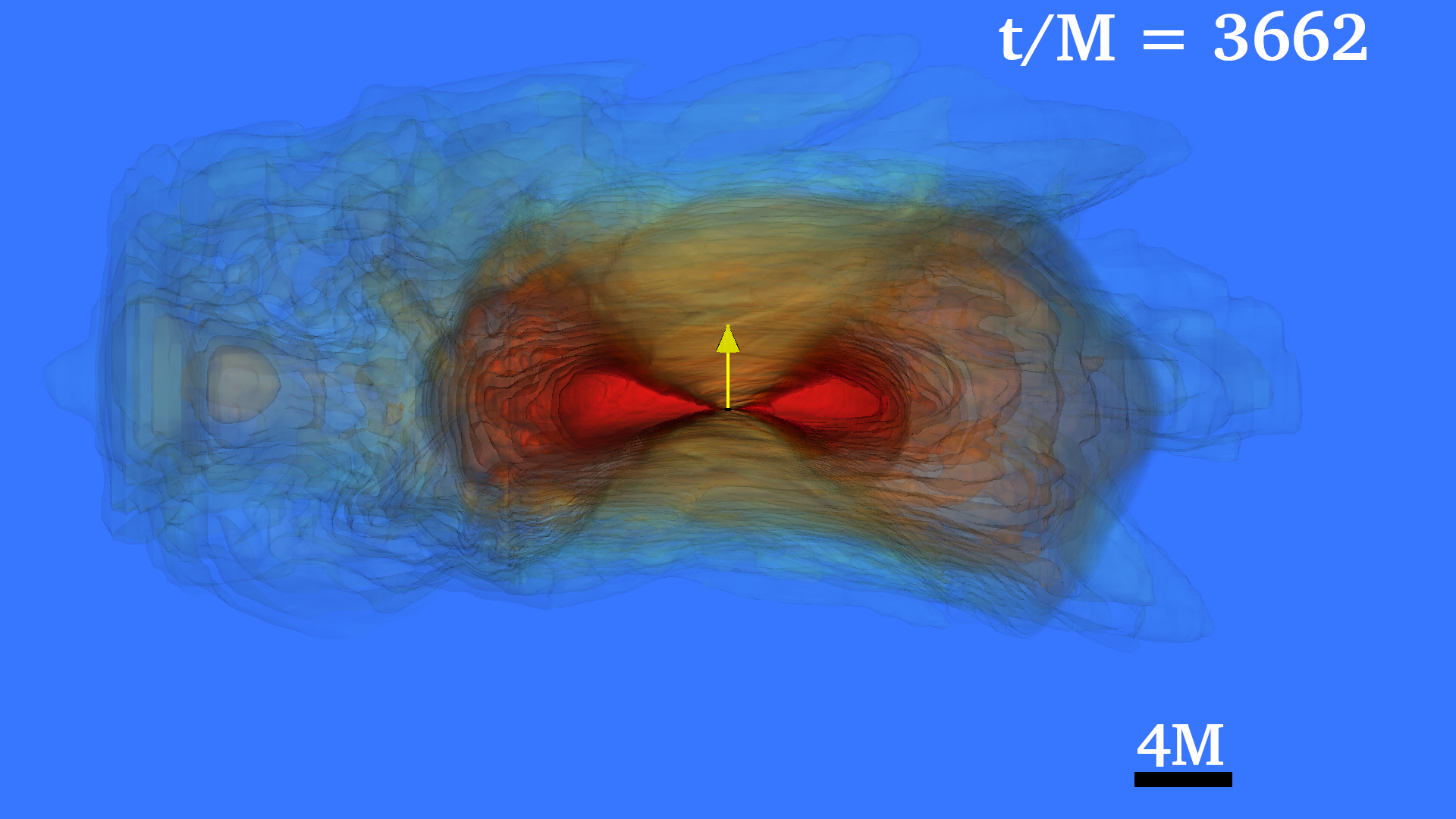}

\rotatebox{90}{\bf\Large $\phantom{mmm}$Tilt: $45^\circ$}
\includegraphics[width=0.95\columnwidth]{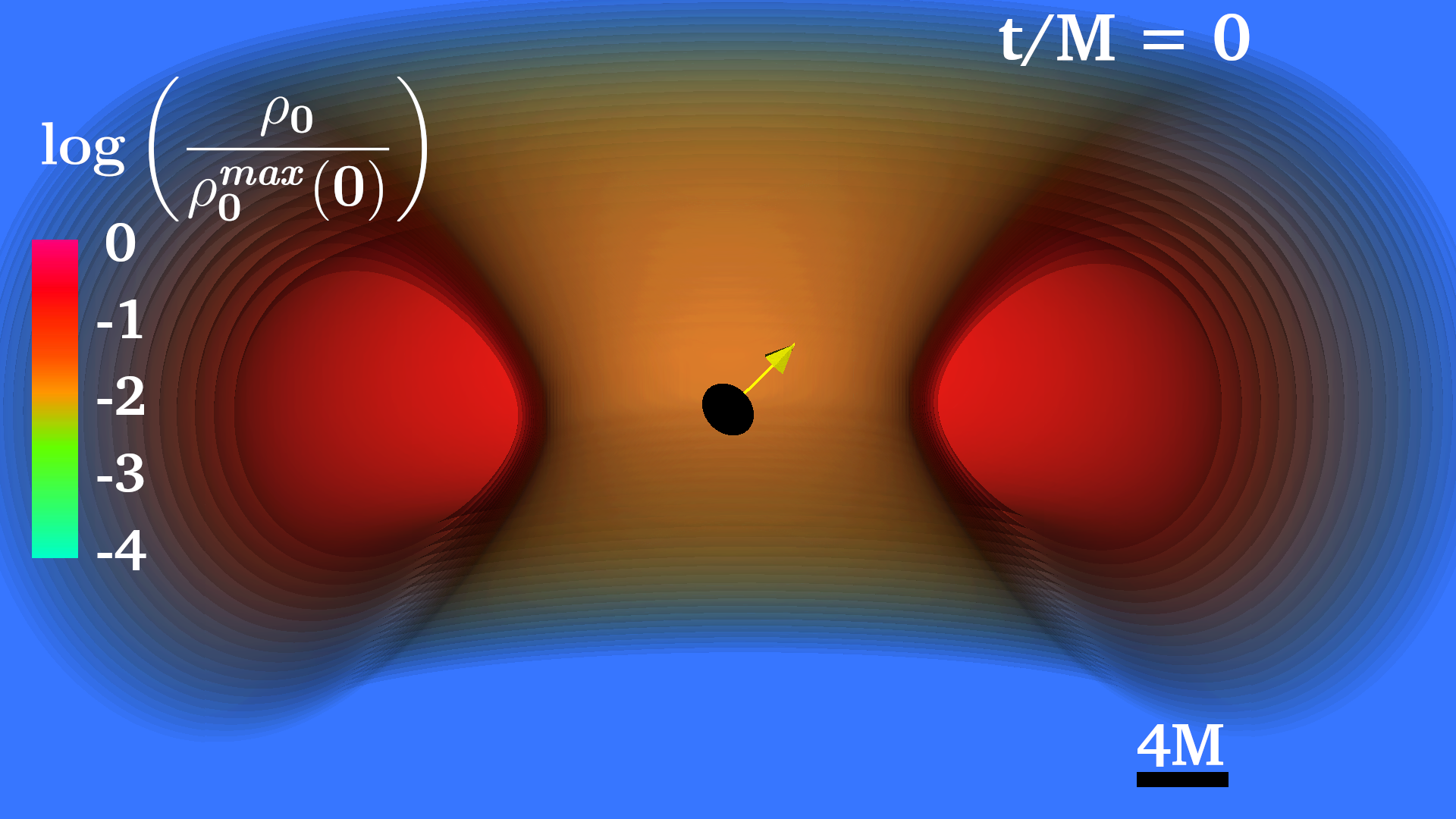}
\includegraphics[width=0.95\columnwidth]{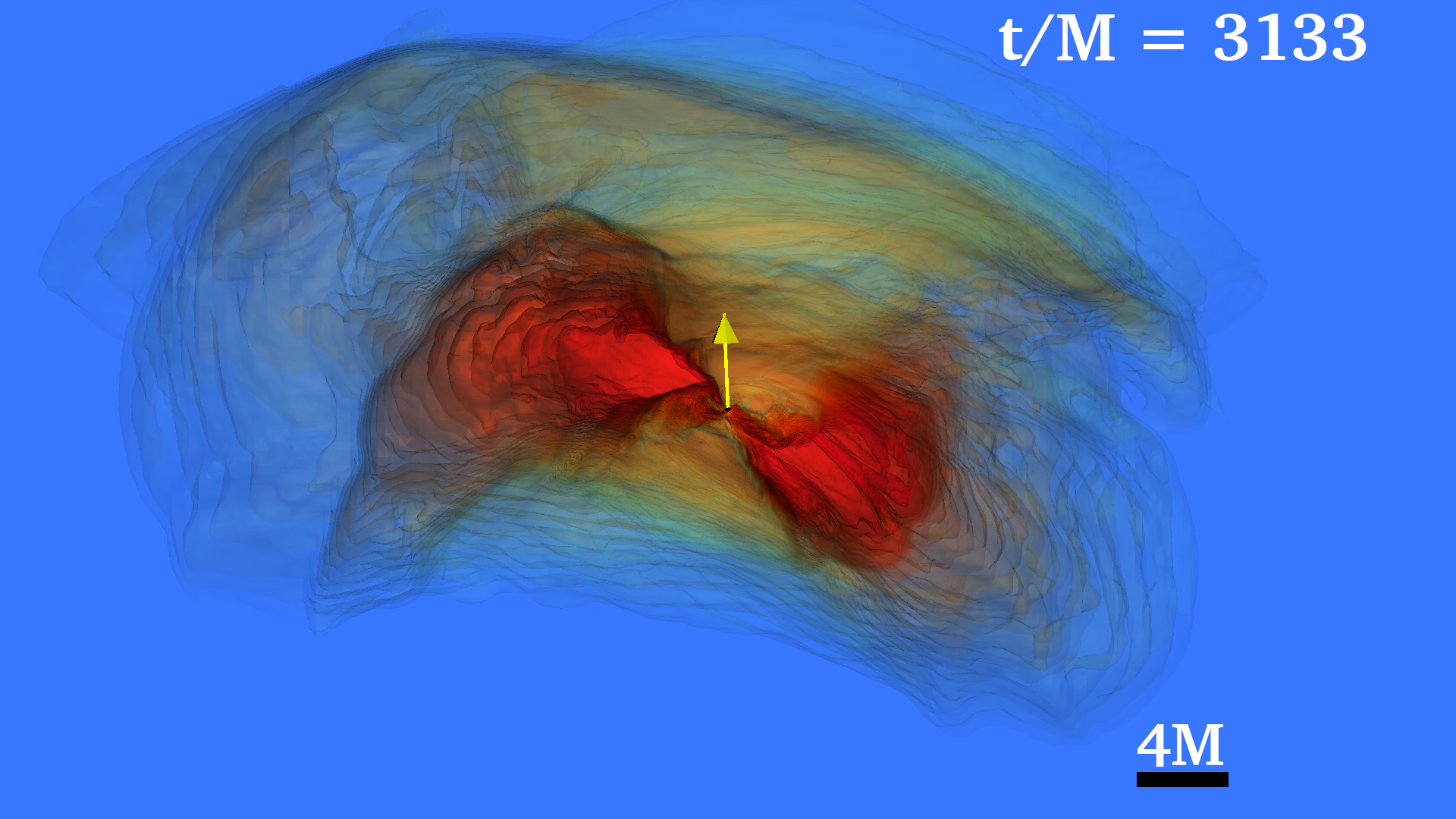}

\rotatebox{90}{\bf\Large $\phantom{mmm}$Tilt: $90^\circ$}
\includegraphics[width=0.95\columnwidth]{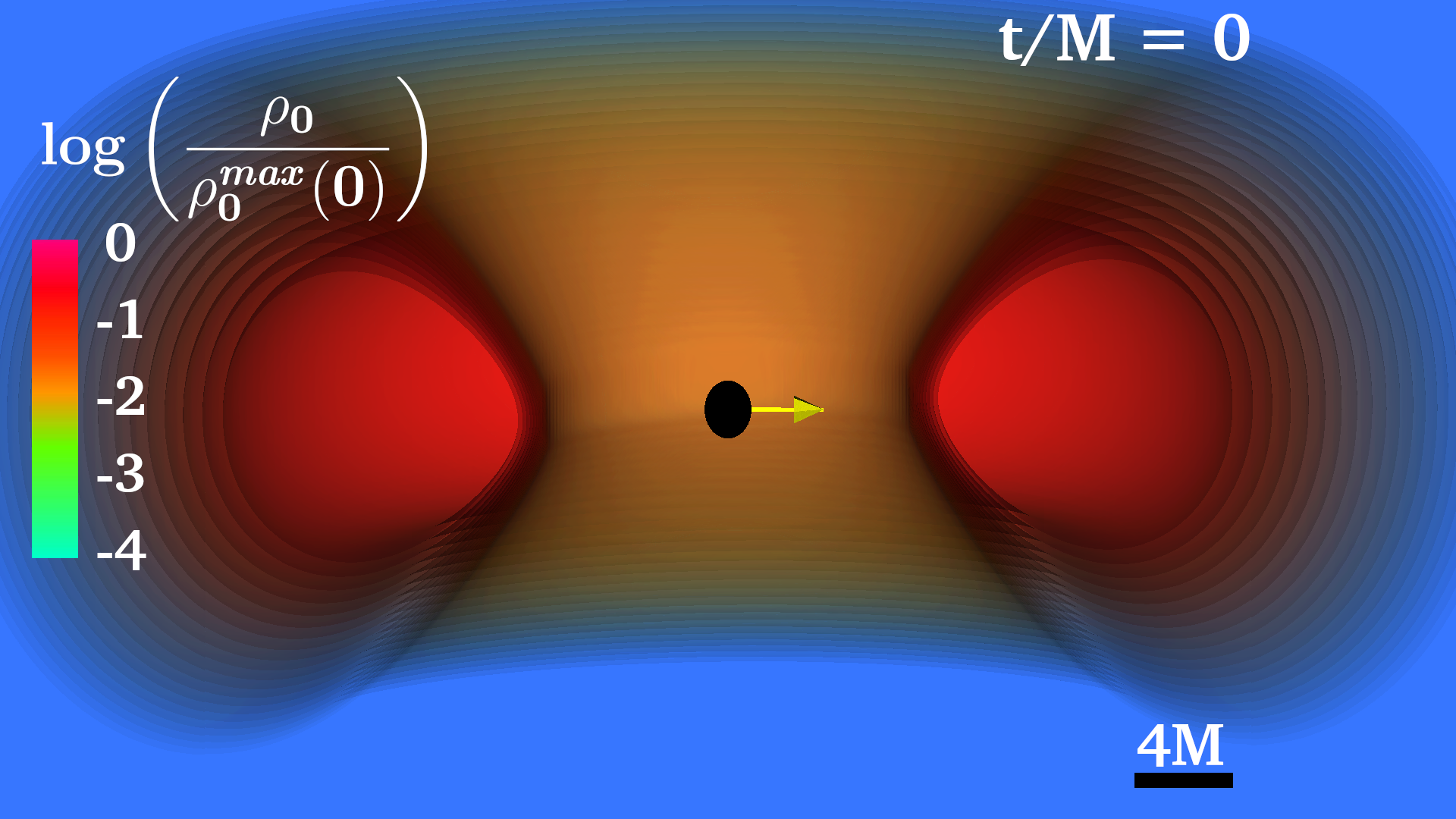}
\includegraphics[width=0.95\columnwidth]{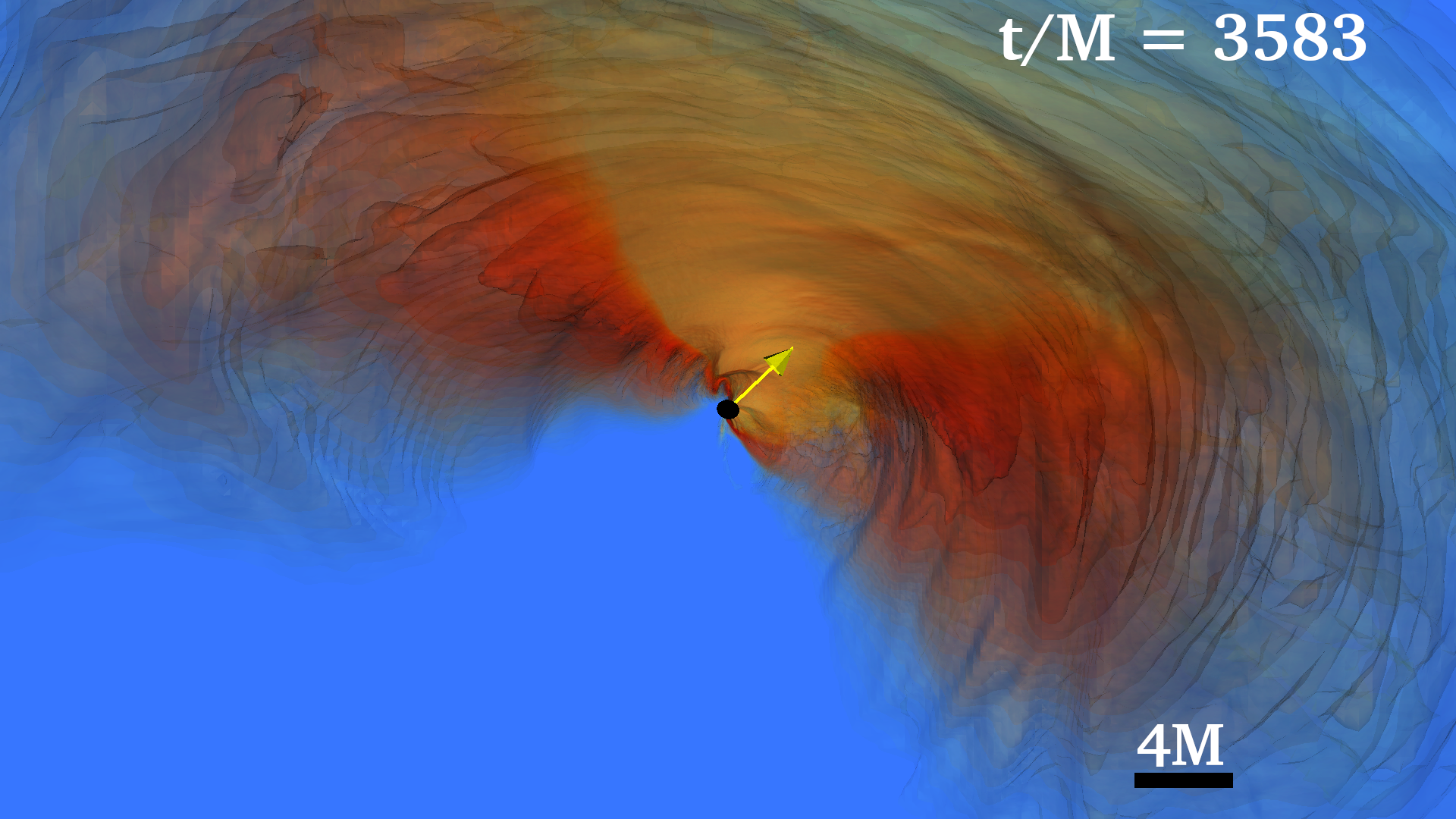}

\rotatebox{90}{\bf\Large $\phantom{mmm}$Tilt: $180^\circ$}
\includegraphics[width=0.95\columnwidth]{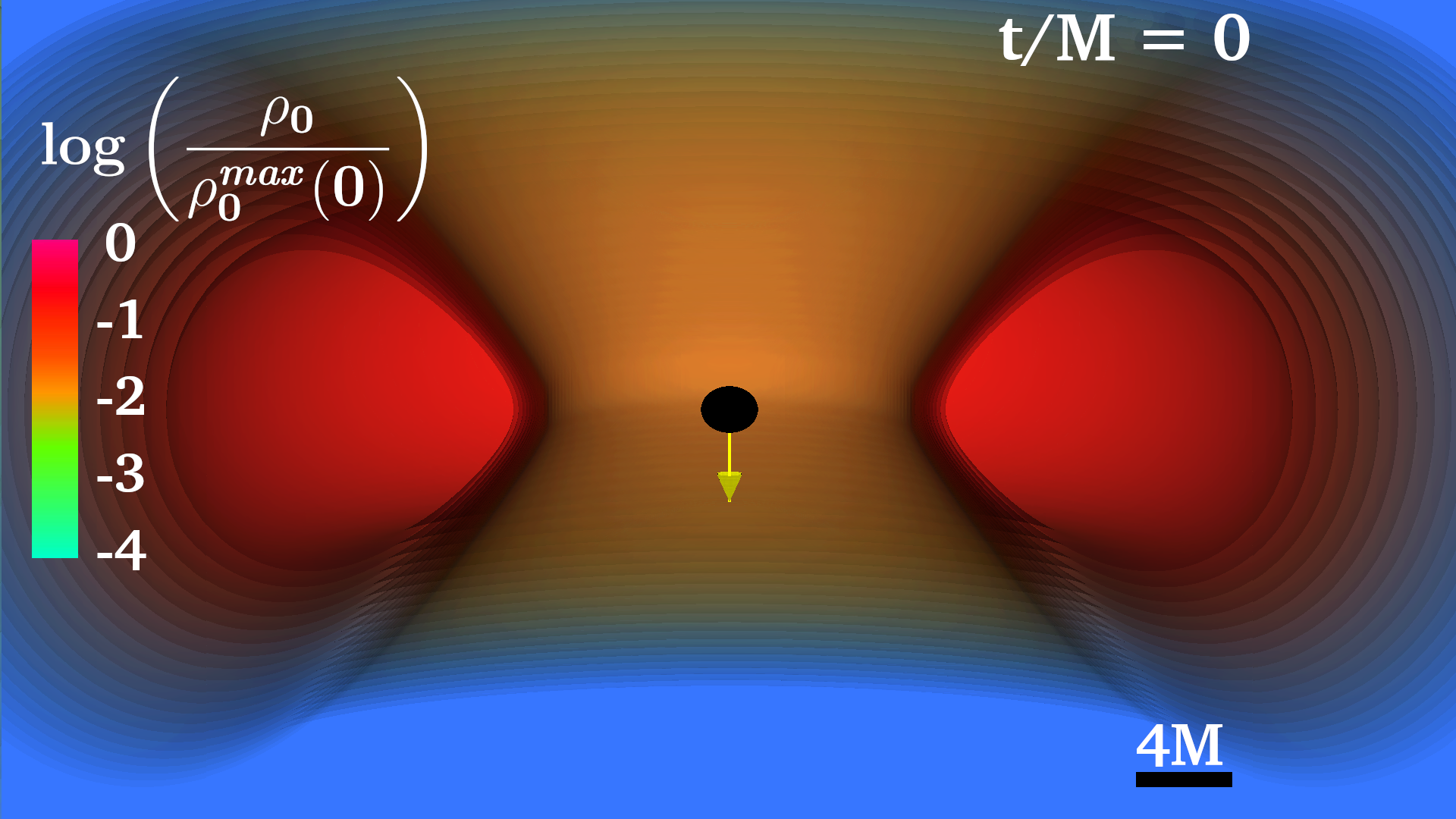}
\includegraphics[width=0.95\columnwidth]{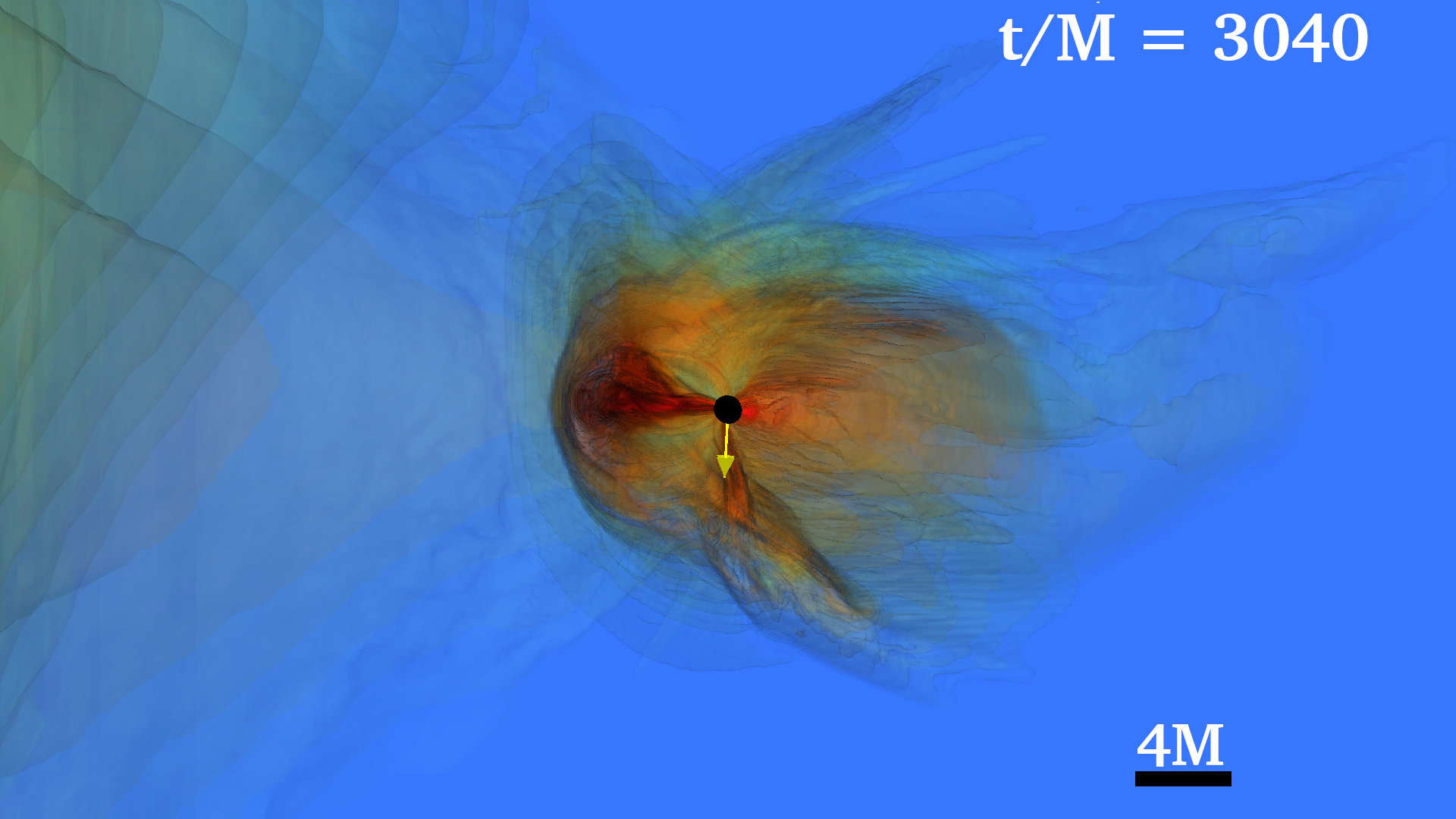}
\caption{Meridional cuts for the initial (left column)  and final (right column) 
state of the rest-mass density for models A1 (first row), 
A2 (second row), A3 (third row), and A4 (fourth row). The direction of the BH 
spin is given by the yellow arrow. The black spheroidal regions denote the 
apparent horizon.}
\label{fig:rho}
\end{center}
\end{figure*}

\begin{figure*}
\begin{center}
\rotatebox{90}{\bf\Large $\phantom{mmm}$Tilt: $0^\circ$ }
\includegraphics[width=0.95\columnwidth]{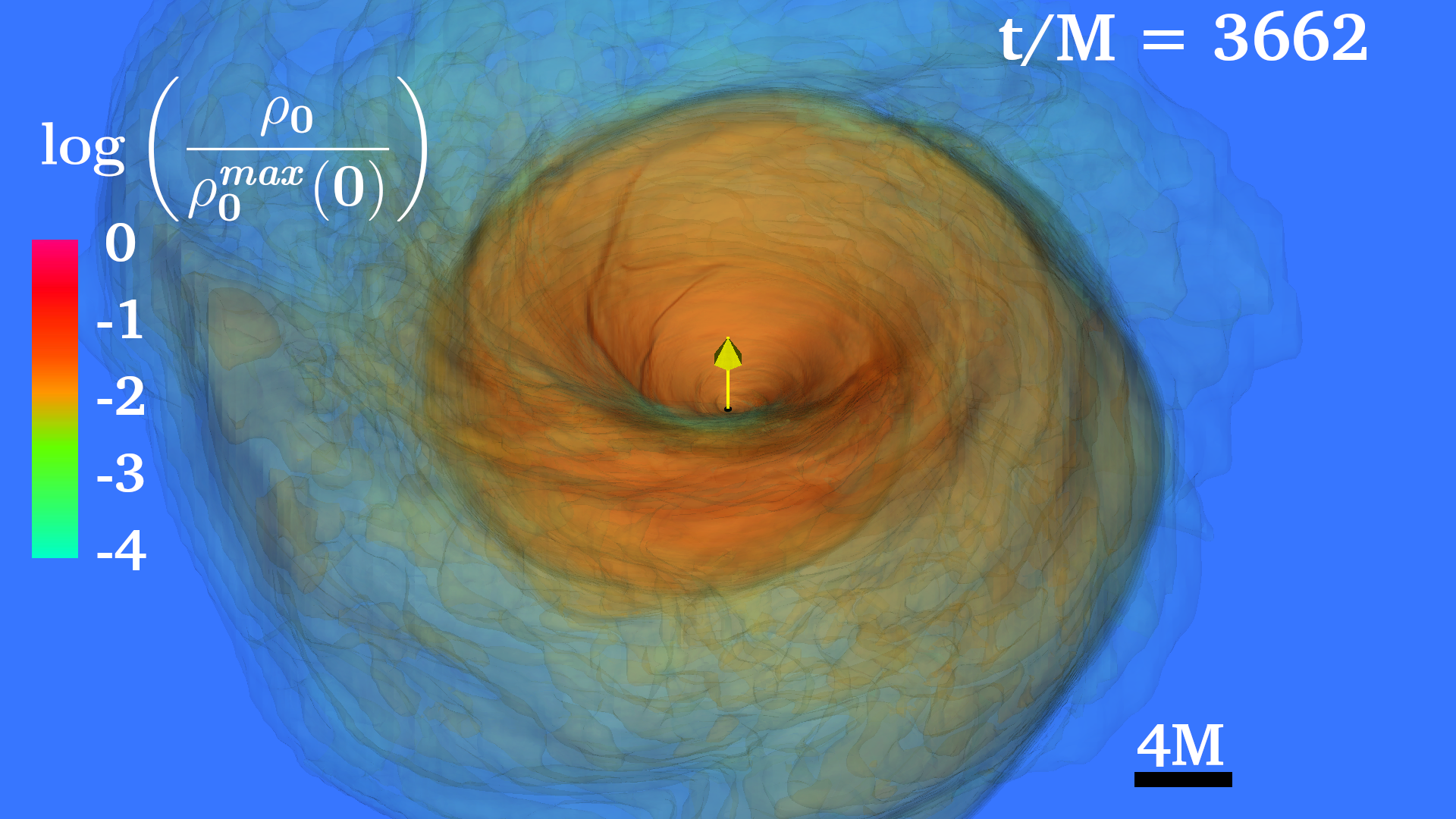}
\includegraphics[width=0.95\columnwidth]{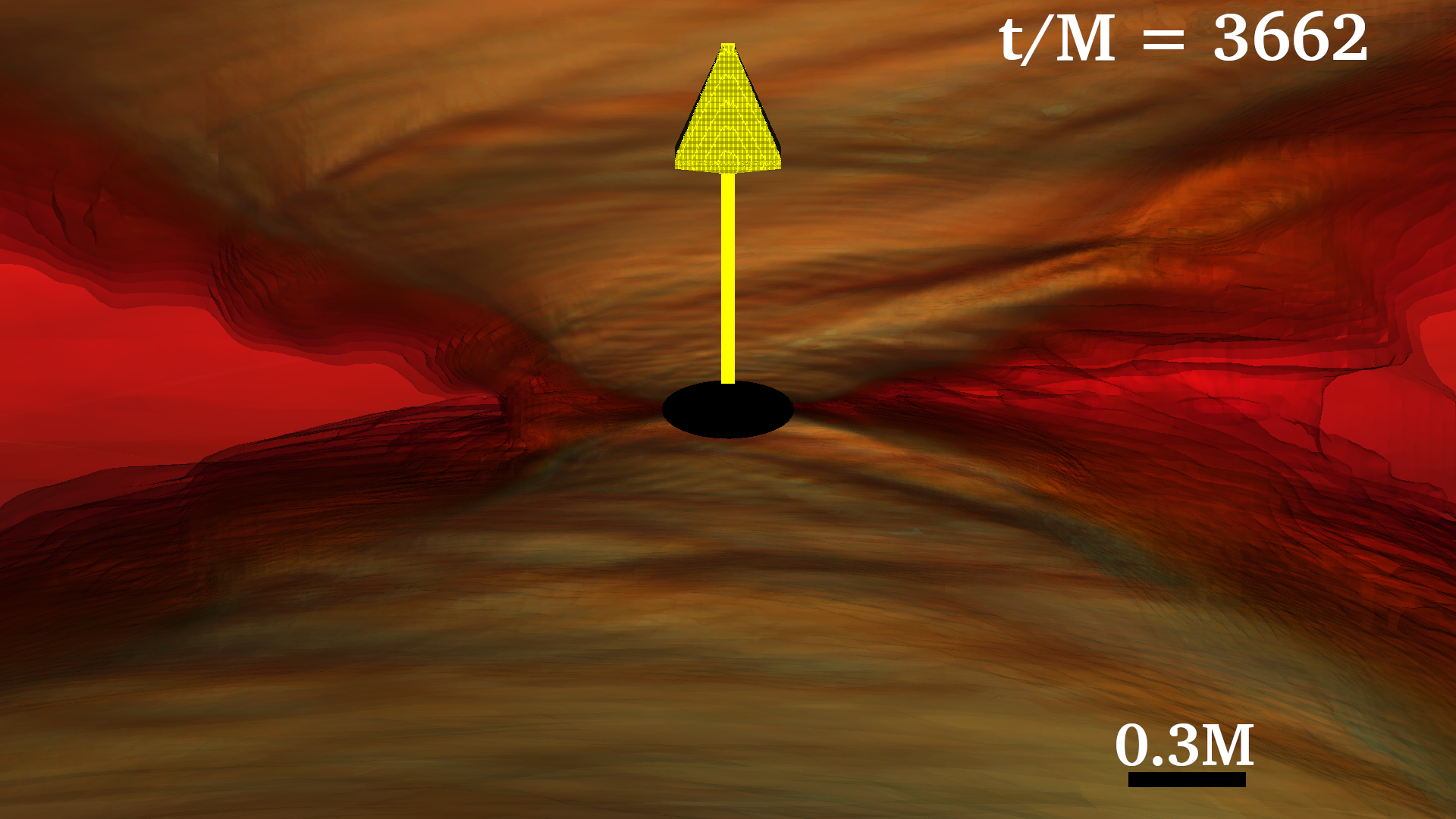}

\rotatebox{90}{\bf\Large $\phantom{mmm}$Tilt: $45^\circ$}
\includegraphics[width=0.95\columnwidth]{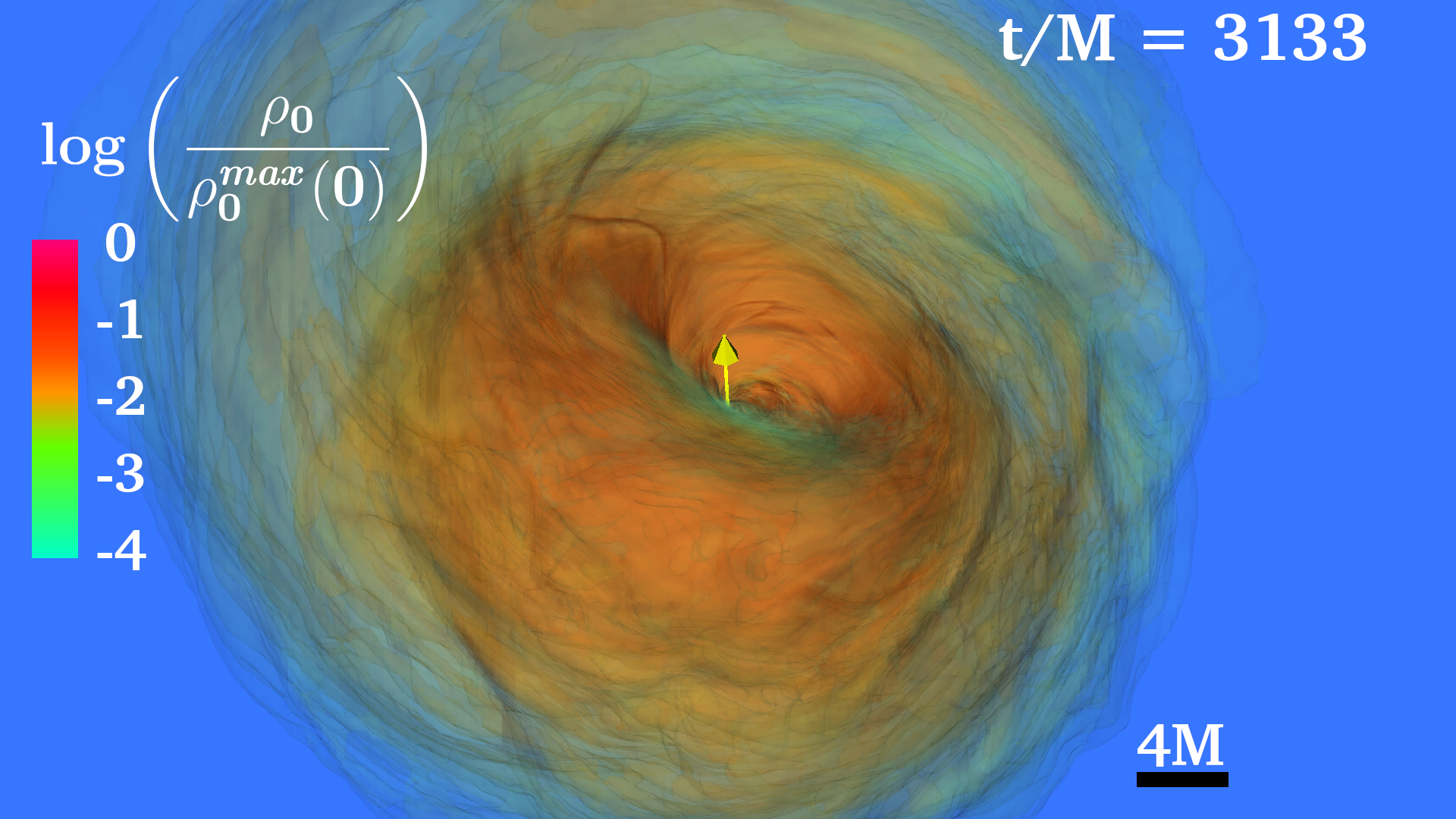}
\includegraphics[width=0.95\columnwidth]{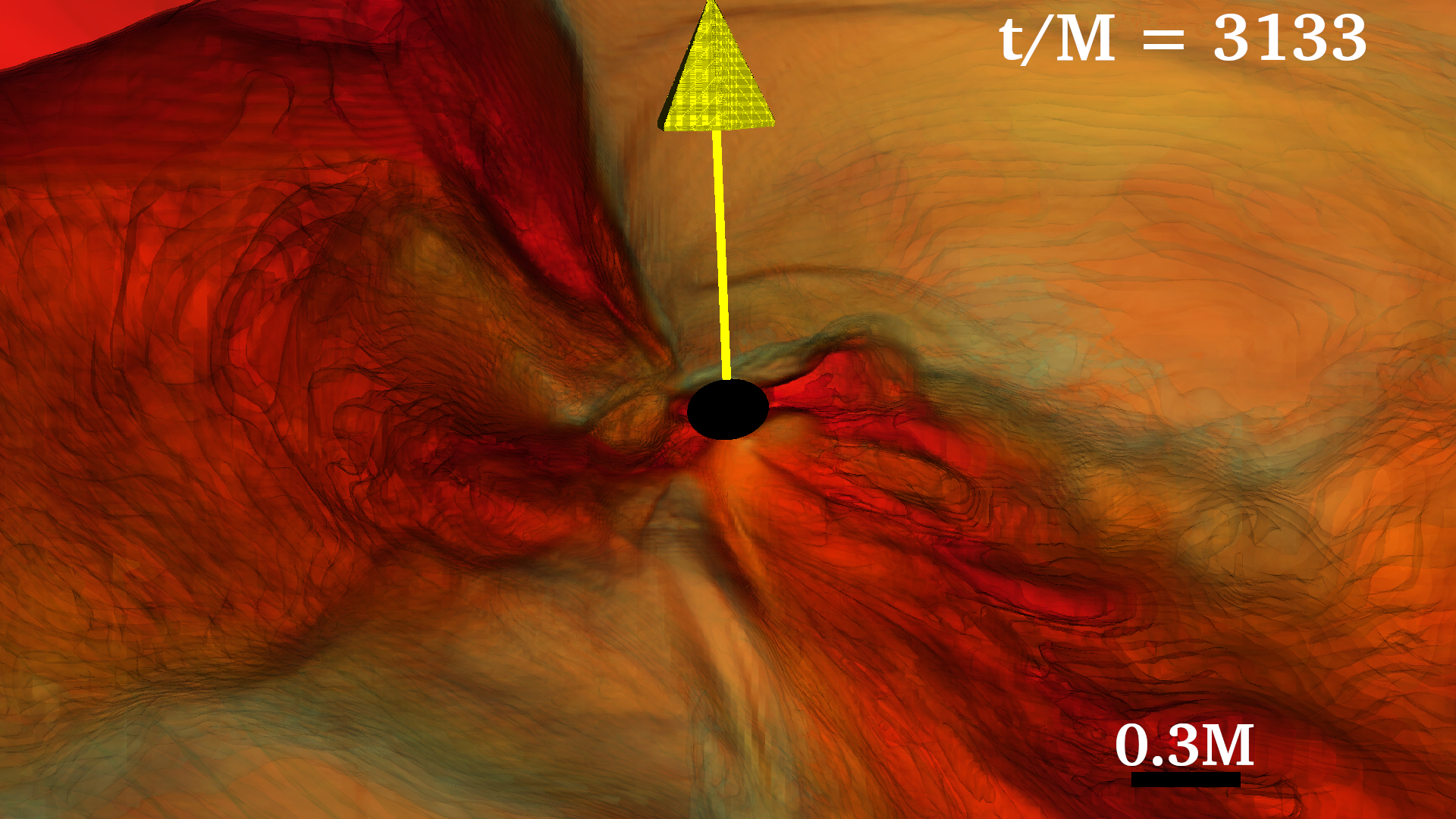}

\rotatebox{90}{\bf\Large $\phantom{mmm}$Tilt: $90^\circ$}
\includegraphics[width=0.95\columnwidth]{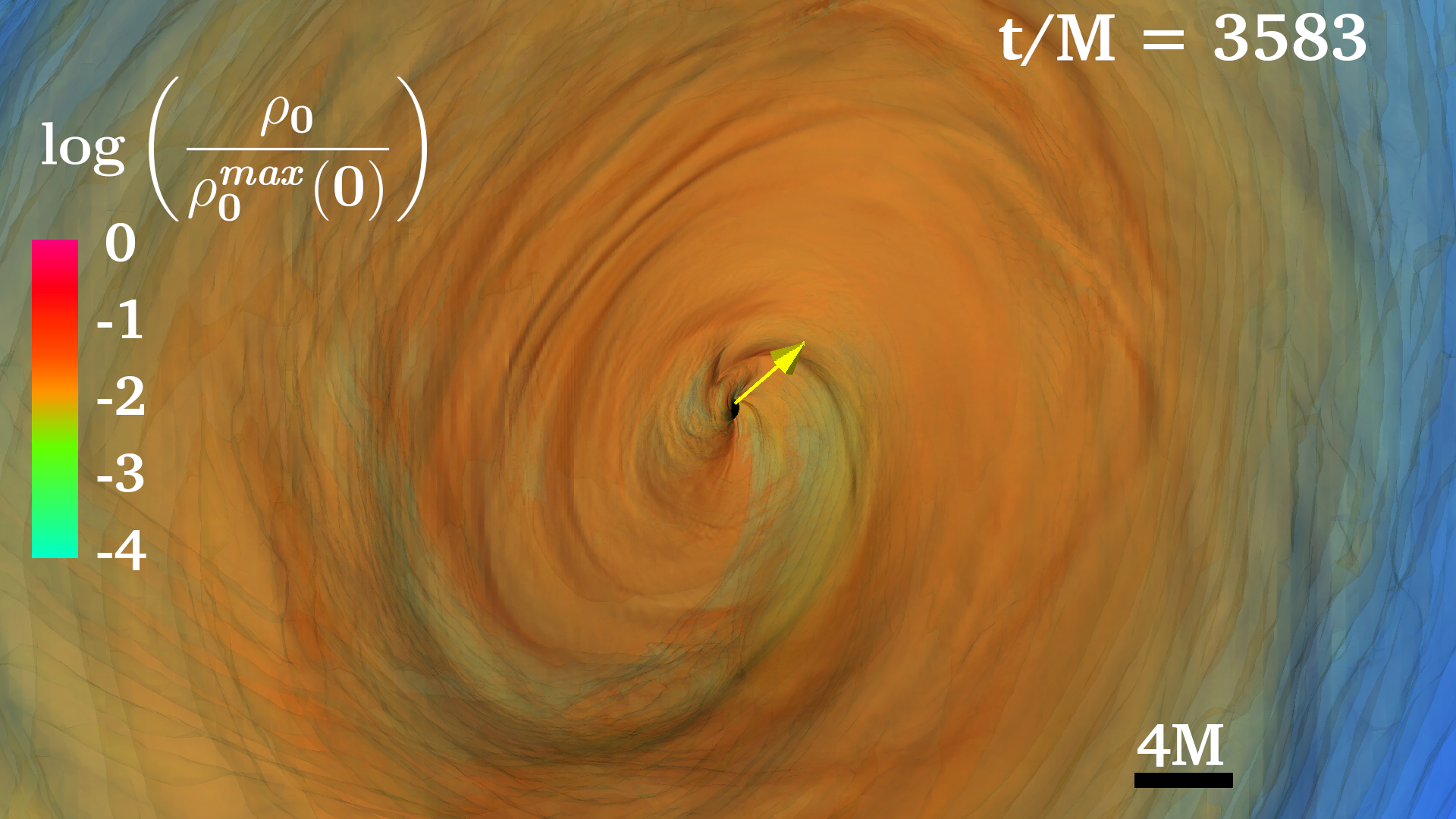}
\includegraphics[width=0.95\columnwidth]{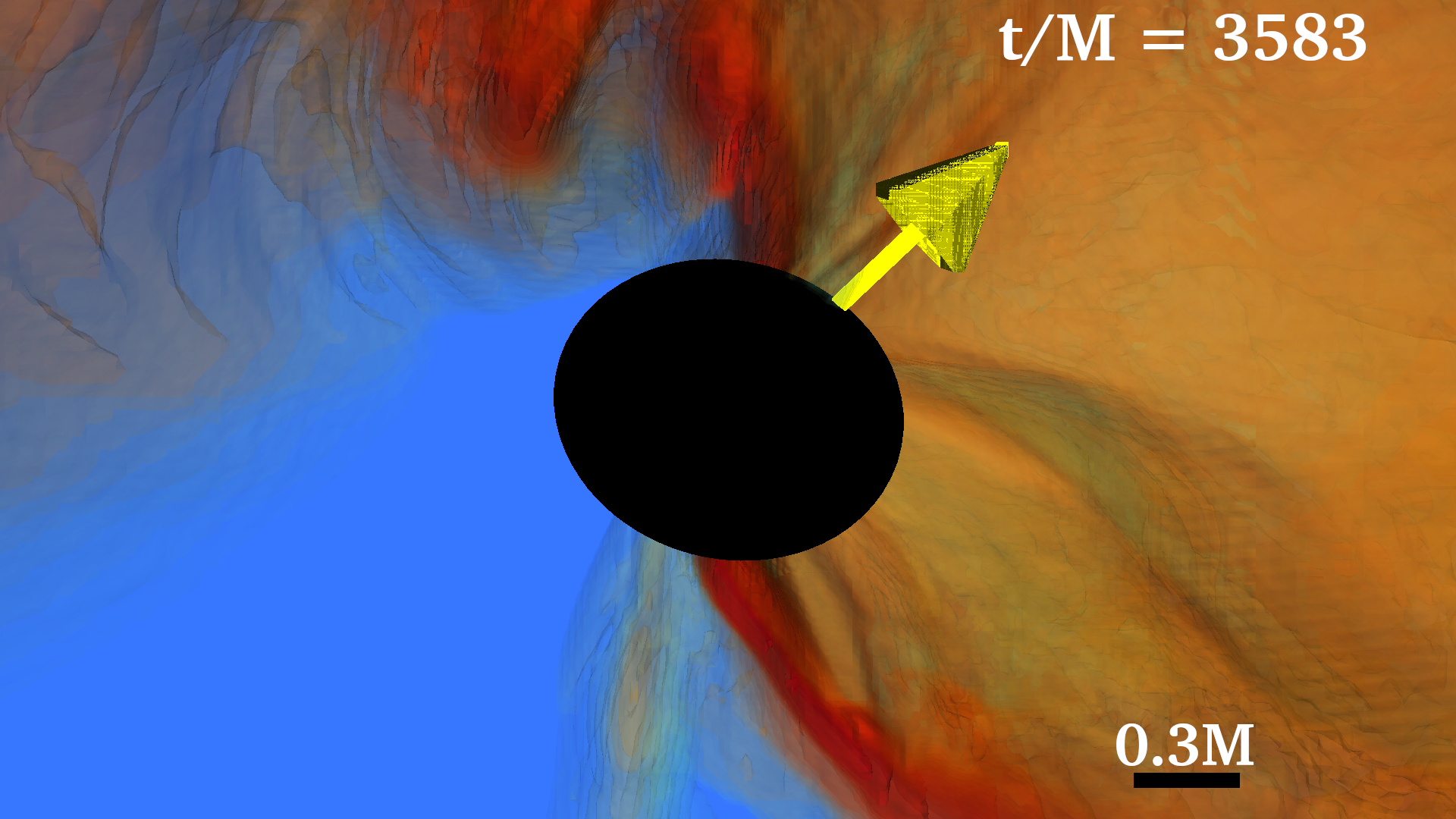}

\rotatebox{90}{\bf\Large $\phantom{mmm}$Tilt: $180^\circ$}
\includegraphics[width=0.95\columnwidth]{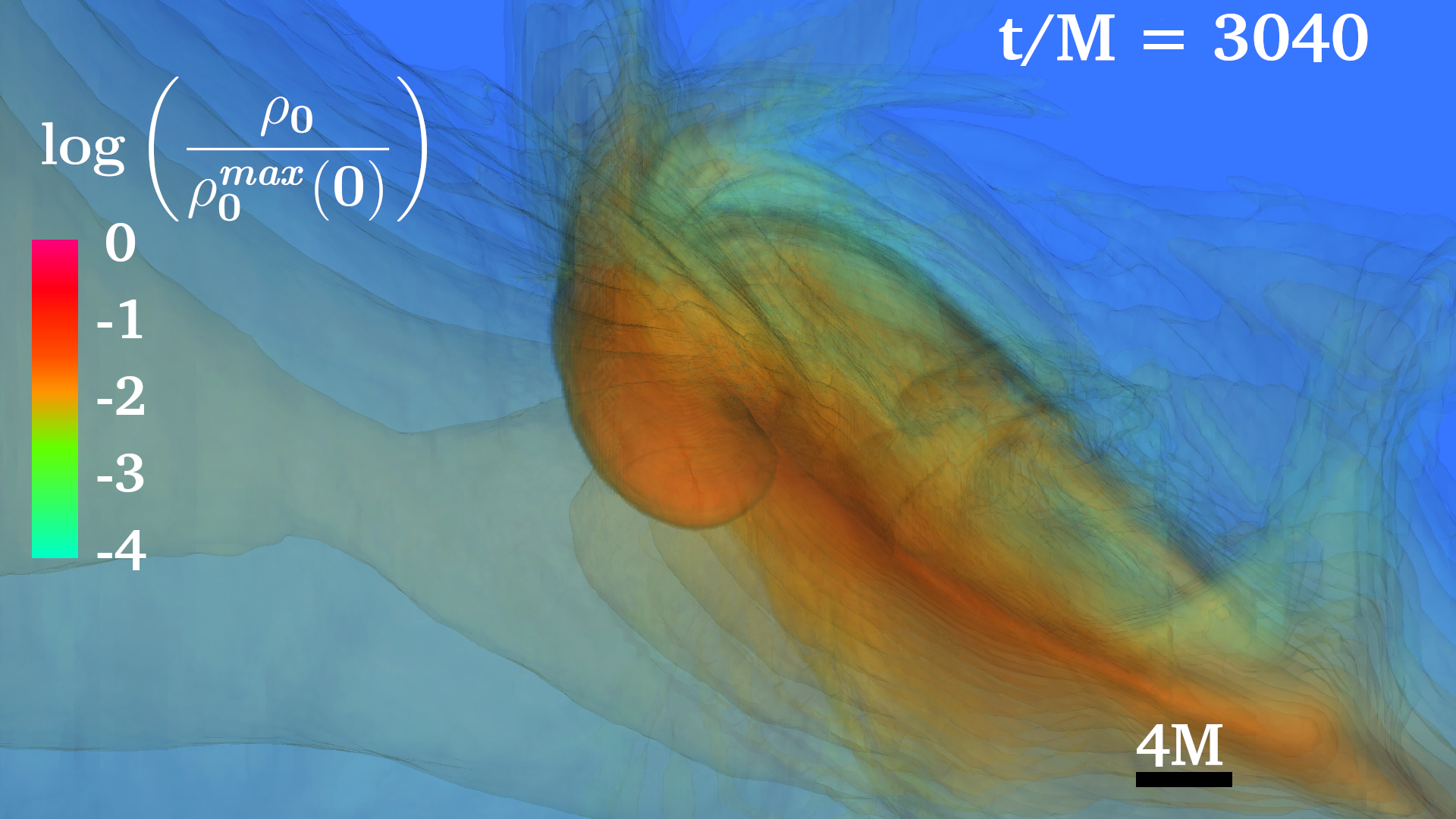}
\includegraphics[width=0.95\columnwidth]{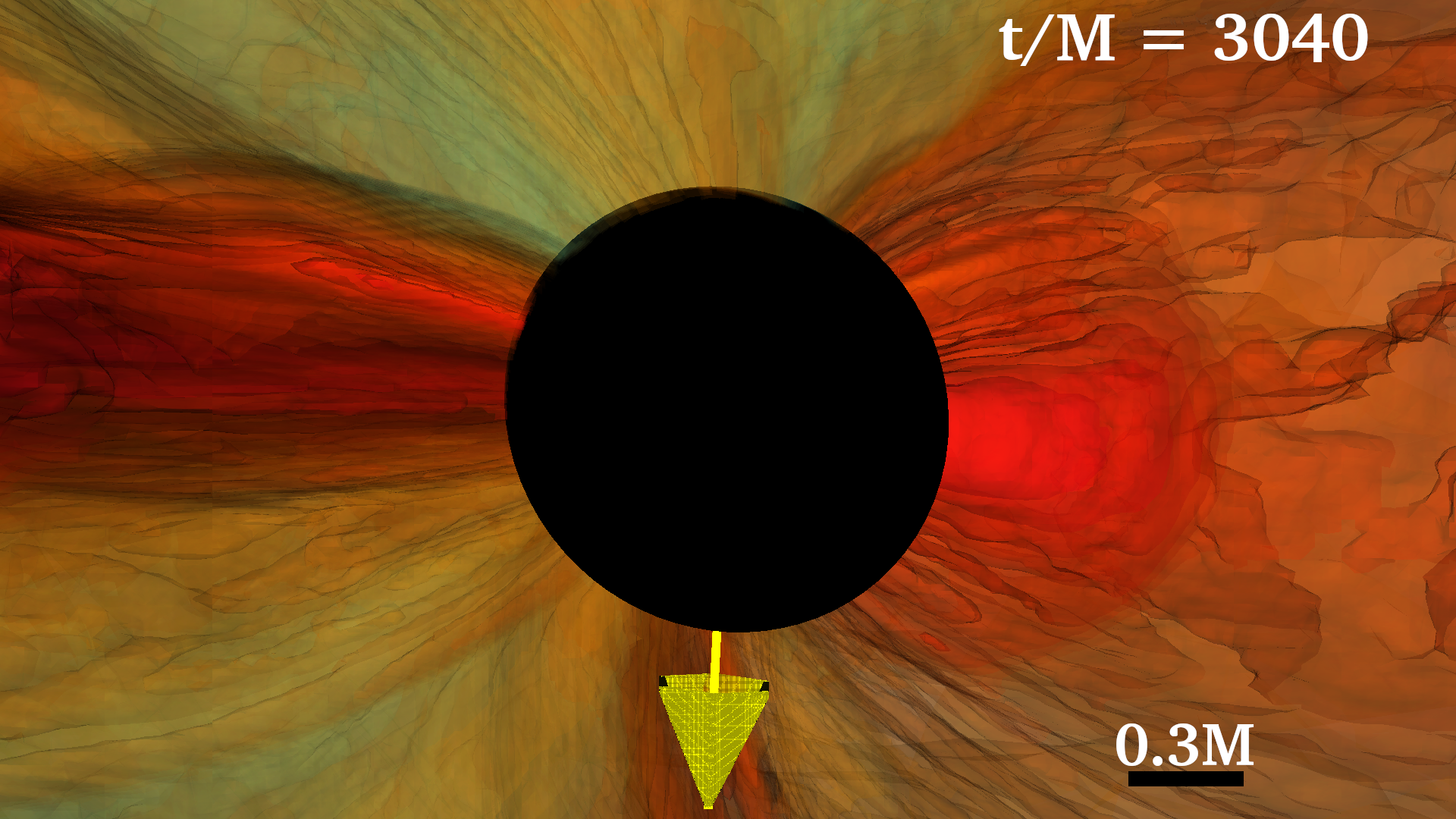}
\caption{Left column full 3d rendering of the disk rest-mass density at the final moment in our
simulations. The right column  zooms in near the BH at the same time as the left column. The
rest-mass densities are plotted for models A1 (first row), A2 (second row), A3 (third row) and
A4 (fourth row). The direction of the BH spin is given by the yellow arrow. The black spheroidal
regions denote the apparent horizon.}
\label{fig:rhozoom}
\end{center}
\end{figure*}

\section{Evolutions}
\label{sec:evol}

The models A1-A4 of self-gravitating BHDs are evolved  using the \illinois
moving-mesh-refinement code that employs the
Baumgarte--Shapiro--Shibata--Nakamura (BSSN) formulation of the Einstein's
equations~\cite{shibnak95,BS} to evolve the spacetime fields.
Outgoing wave-like boundary conditions are applied to all BSSN variables, which
are evolved using the equations of motion (9)-(13) in~\cite{Etienne:2007jg},
along with the $1+$log time slicing for the lapse $\alpha$, and the
``Gamma--freezing" condition for the shift $\beta^i$, cast in first-order form
(see~Eq.~(2)-(4) in~\cite{Etienne:2007jg}).  Time integration is performed via
the method of lines using a fourth-order accurate Runge-Kutta integration
scheme with a Courant-Friedrichs-Lewy factor set to $0.36048$.  Spatial
derivatives are computed with fourth-order, centered finite differences, except
on shift advection terms, where we employ fourth-order upwind differencing.  We
use the Carpet infrastructure~\cite{Carpet,carpetweb} to implement moving-box
adaptive mesh refinement, and add fifth-order Kreiss-Oliger
dissipation~\cite{goddard06} to spacetime and gauge field variables.  For
numerical stability, we set the damping parameter $\eta$ appearing in the shift
condition to $\eta\approx 26.6/M$.  For further stability we modify the
equation of motion of the conformal factor $\phi$ by adding a
constraint-damping term (see Eq.~(19) in~\cite{DMSB}) which damps the
Hamiltonian constraint. We set the constraint damping parameter to $c_H=0.08$
(see also \cite{Raithel2022}).  

High resolution, shock-capturing methods \cite{Etienne:2011re, Etienne:2010ui}
are used for the equations of hydrodynamics, which are written in conservative
form.  The primitive, hydrodynamic matter variables are the rest-mass density,
$\GR_0$, the pressure $P$ and the coordinate three velocity $v^i=u^i/u^0$. The
stress energy tensor is $T_{\GA\GB}=\GR_0 hu_\GA u_\GB+P g_{\GA\GB}$.  For the
EOS we use the ideal gas
$\Gamma$-law $P=(\Gamma-1)\GR_0\GE$ with $\Gamma=4/3$, and $\GE$ the specific
internal energy.  The grid hierarchy used in our simulations is summarized in
Table \ref{tab:evol}. It consists of a set of 13 nested mesh refinement boxes
centered on the BH apparent horizon. The computational domain is
$[-4000\mbh,4000\mbh]^3$.  The half-side length of the finest box 
has $\Delta x_{\rm min}=50\mbh/2^{12} = 0.0122\mbh$. 
Note that the ADM mass is $M\approx 1.2\mbh -1.3\mbh$ depending on the model.  
In our simulations we do not assume any symmetry. The extremely high resolution
used is necessary in order to capture accurately the dynamics of the highly
spinning BHs.

\subsection{Global structure}

The overall evolution of models A1-A4 can be seen in Figs. \ref{fig:rho} and
\ref{fig:rhozoom}. At $t=0$ (left column of Fig. \ref{fig:rho}) the disks have
very similar geometries (see also Fig. \ref{fig:rhoj}) while the BHs have the
same mass and similar spin magnitudes. Thus the main difference in our cases is
the BH tilt angle, which results in distinct behaviors for the 4 models. Note
that in Figs. \ref{fig:rho} and \ref{fig:rhozoom} the magnitude of the BH spin
vector is not to scale. 
Also the shrinkage of the BH and the disk sizes in the right column of Fig. 
\ref{fig:rho}, and the left column of Fig. \ref{fig:rhozoom} are due to gauge
effects arising from differences between the initial data and the evolution
gauge choices.
On the right column of Fig. \ref{fig:rho} we depict a
meridional cut at the final moment in our evolutions. For the aligned case (top
row) the BH preserves its spin orientiation and magnitude and the
disk retains its broad characteristics. The one-arm instability fully develops,
but the induced BH orbit remains bounded. On the other hand, the antialigned case
(bottom row) after a certain time becomes largely unstable, with the disk losing
its initial structure and exhibiting massive mass accretion. The BH acquires a
kick velocity that results in an unbound orbit (keeps drifting away until 
the end of our simulations). Although the BH spin orientiation is preserved, 
its magnitude  is significantly reduced due to accretion. 

For the misaligned cases (second and third rows), we observe the combined effects of (i) BH
precession, (ii) disk precession and warping around the BH, (iii) development of
the PPI, (iv) acquisition of a small BH kick velocity, (v) significant
gravitational wave emission of various modes beyond the $\ell=2$, $m=2$ which
are as strong as the $(2,2)$ mode, and (vi) in the A3 case ($90^\circ$ initial
tilt), we observe an overall broad alignment of the disk with the BH spin (third
row in Fig. \ref{fig:rho}, right column).  This alignment is not associated
with the  BP effect, which requires a viscosity mechanism absent
in our simulations. In addition, the alignment in our case is global, i.e.
the whole disk rotates like a solid body, instead of the alignment of only the
inner regions of the disk typical of the BP picture. In fact, from the third
row, right column of Fig. \ref{fig:rhozoom}, where the two streams onto the BH
are apparent, we confirm that there is no such alignment in the inner regions
of the disk. Our results are reminiscent of the behavior described in
\cite{Fragile2005,Fragile2007b} and referred as ``plunging streams''.  The
additional complication in our case though is that the BH-disk spacetime is
dynamical and responds to the motion of the disk. As in
\cite{Fragile2005,Fragile2007b} the plunging streams enter the BH above and
below its symmetry plane from almost antipodal points due to strong
differential precession and the nonspherical nature of the spacetime. For a
Kerr BH (which is very close to the BHD spacetimes close to the horizon)
orbital stability strongly depends on the inclination of the orbit, with the
unstable region being larger for increasing inclination. Also the value of
$r_{\rm ISCO}$ is larger for larger inclinations
\cite{Hughes:2001jr,Fragile2007b}. For the A3 case we observe the largest BH
kick velocity which is $\sim 2\ \rm km/s$. For model A2 ($45^\circ$ initial
tilt angle) we could not evolve beyond $t\approx 3133 M$ because the BH was
spun up to maximal spin. At that point both the BH and the
disk experience a tilt by $\sim 45^\circ$ with respect to their initial
orientation, but in opposite directions (see second row, right column in Fig.
\ref{fig:rho}) Similar to case A3 and \cite{Fragile2005,Fragile2007b}, we
observe two plunging streams in opposite directions entering the BH above and
below its symmetry plane.  The warping of the disk around the BH for both cases A2
and A3 is significant (see Fig. \ref{fig:rhozoom} second and third row).

\setlength{\tabcolsep}{5pt}                                                      
\begin{table}                                                             
\caption{Mode growth, pattern speed, and corotating radius for the $m=1$ mode.}                       
\label{tab:mg}                                                     
\begin{tabular}{cccccccccccccc}                                                    
\hline\hline                                                              
Model & ${\rm Im}(\GO_1)/\Omega_c$ & $\Omega_{p,1}/\Omega_c$ & $r_{\rm cr}/r_{\rm c}$  \\ \hline\hline
A1    & $0.318$   & $0.748$  &  $1.17$   \\ \hline
A2    & $0.177$   & $0.748$  &  $1.17$   \\ \hline
A3    & $0.177$   & $0.637$  &  $1.24$   \\ \hline
A4    & $0.227$   & $0.812$  &  $1.12$   \\ \hline
\end{tabular}                                                               
\end{table}                                                                   
%

\subsection{Mode growth and angular momentum transport}
\label{sec:mg}

According to previous studies, both Newtonian and general relativistic, we
expect all our models to be dynamically unstable to the one-arm ($m=1$) spiral-shape
instability. In the general relativistic simulations of
\cite{Korobkin2011,Mewes2015,Mewes2016} it was concluded that if the mass of
the disk is larger than $\gtrsim 4\%$ of the mass of the BH a fixed background
spacetime cannot fully capture the dynamics of the system.  In particular in
order to accurately describe the dynamical gravitational interaction between a
time varying BH (in position, mass and spin), as well as a time
varying massive disk, simulations in a non-fixed background spacetime are
necessary, as we perform here.

\begin{figure}
\begin{center}
\includegraphics[width=0.99\columnwidth]{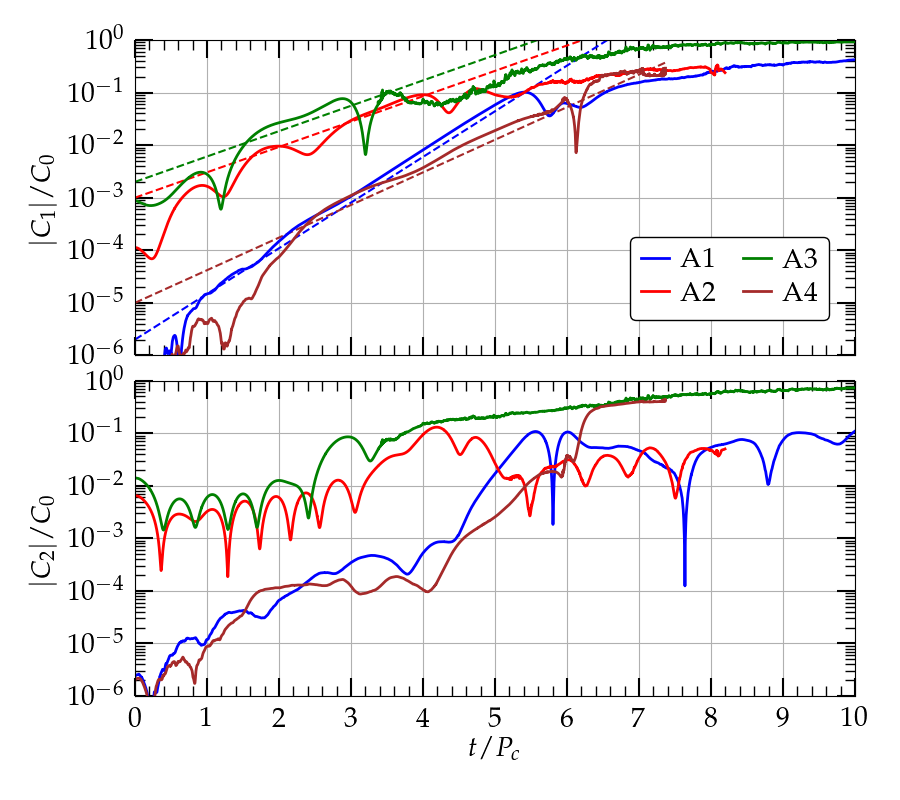}
\caption{Growth of the $m=1$ (top panel) and the $m=2$ (bottom panel) modes. }
\label{fig:cm}
\end{center}
\end{figure}

\begin{figure}
\begin{center}
\includegraphics[width=0.99\columnwidth]{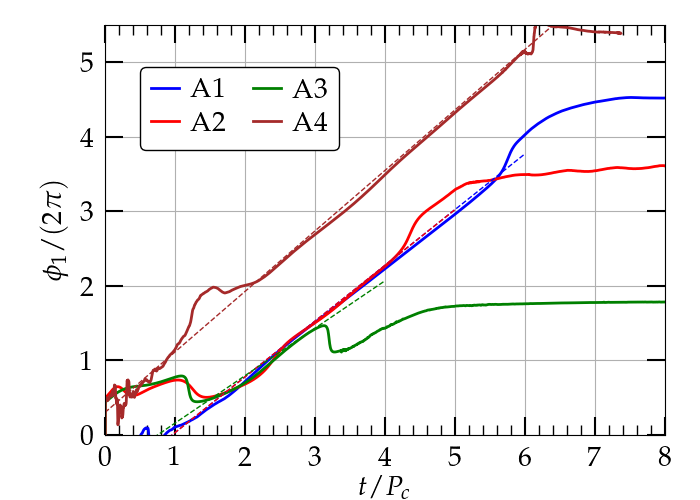}
\caption{Phase angle $\GP_1$ of the mode $m=1$ for models A1-A4.}
\label{fig:phi1}
\end{center}
\end{figure}

To quantify the growth of various unstable density modes we evaluate
the parameters \cite{Paschalidis:2015mla,Wessel:2020hvu} 
\be
C_m =  \int_{r>r_{\rm ah}} \GR_0 u^t \sqrt{-g} e^{im\GP} d^3x  ,
\label{eq:cm}
\ee
where $g$ is the determinant of the spacetime metric and $\GP=\tan^{-1}(y/x)$
the azimuthal angle.  The volume integral is performed outside the apparent
horizon of the BH and the mode amplitude is denoted by the normalized quantity
$C_m/C_0$, where $C_0=M_0$ the rest mass of the disk.  The pattern speed of an
azimuthal mode $m$ is defined as \cite{Williams1987,Woodward1994} 
\be
\Omega_{p,m} = \frac{1}{m} \frac{d\GP_m}{dt} ,
\label{eq:modeome}
\ee
with the phase angle $\GP_m$ being
\be
\GP_m = \tan^{-1}\left(\frac{{\rm Im}(C_m)}{{\rm Re}(C_m)}\right) .
\label{eq:modephi}
\ee
In other words the pattern speed of any mode is proportional to the slope of the 
curve $\GP_m(t)$ with the proportionality constant being $1/m$.

\begin{figure*}
\begin{center}
\includegraphics[width=2.0\columnwidth]{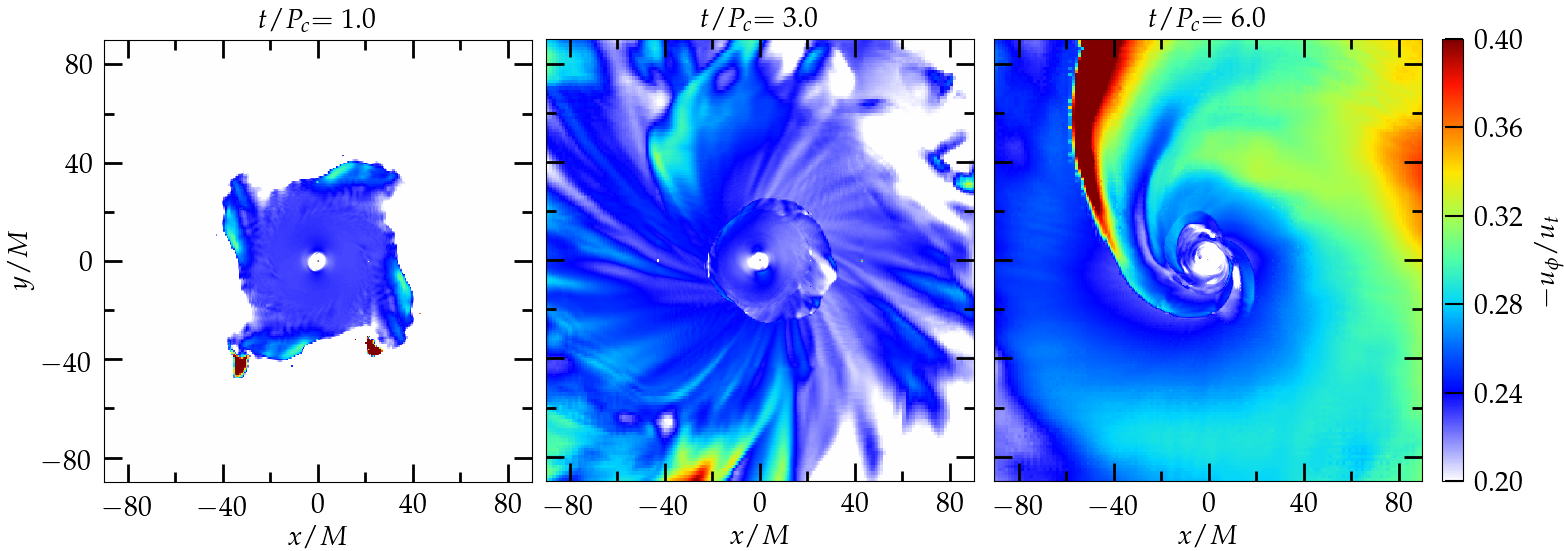}
\caption{Snapshots at three different times of the specific angular momentum 
$\ell=-u_\GP/u_t$ for case A2. To convert to $t/M$ multiply by $373$ (see Table \ref{tab:id}).}
\label{fig:ell}
\end{center}
\end{figure*}

\begin{figure}
\begin{center}
\includegraphics[scale=0.4]{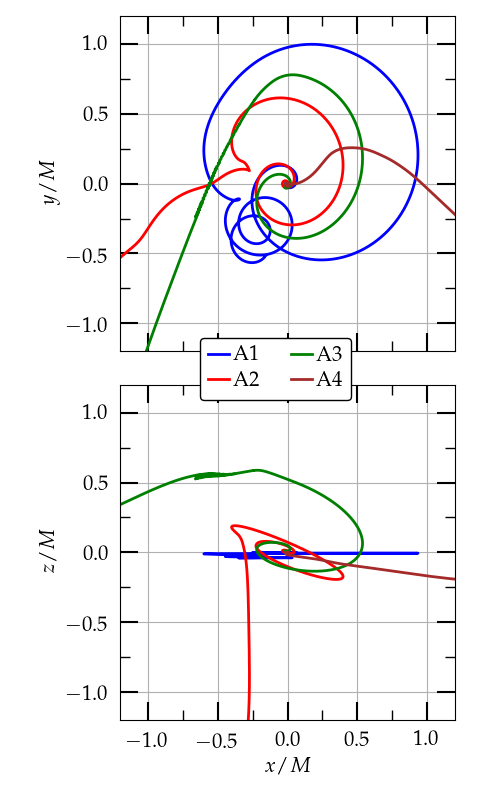}
\caption{BH trajectory on the xy and xz planes.}
\label{fig:BHpos}
\end{center}
\end{figure}

As we discussed in the Introduction, the PPI manifests itself when a
perturbation which is traveling backwards relative to the fluid at the inner
edge, and therefore has $\Omega_{p,m}<\Omega$, exchanges energy and angular
momentum with a perturbation which is traveling forwards relative to the fluid
at the outer edge and therefore has $\Omega_{p,m}>\Omega$. The radius $r_{\rm
cr}$ where the interaction happens is called the corotation radius and
satisfies  $\Omega_{p,m}=\Omega(r_{\rm cr})$.

In Fig. \ref{fig:cm} we plot the $m=1$ (top panel) and $m=2$ (bottom panel)
mode growths for all cases A1 (aligned, blue line), A2 ($45^\circ$ red line) A3
($90^\circ$ green line), and A4 ($180^\circ$ brown line). The most prominent
feature of this plot is the fact that for both modes, $(C_m/C_0)(t=0)$ for the
tilted cases (A2, A3) are much larger than the ones of A1, A4\footnote{
For models A1, A4, at $t=0$ slight deviations from $C_m\equiv 0$, $m\geq 1$,
are due to numerical error, as the disks are constructed to be strictly 
axisymmetric in spherical polar coordinates and then interpolated onto a 
Cartesian grid.}. 
In fact for models A2 and A3 $(C_2/C_0)(t=0) \sim
O(10^{-2}) $ is ten to a hundred times larger than $(C_1/C_0)(t=0)$ and initially
slightly decreases while the latter steadily grows in an exponential manner.
When $C_1/C_0$ reaches values $\sim O(10^{-2}) $  then the $m=2$ mode grows in
a similar manner.  In other words the $m=1$ mode drives the growth of the $m=2$,
something that is also seen in the aligned and antialigned cases (A1, A4). The
fact that in the tilted cases at $t=0$ the $m=1$ mode amplitude is already nonzero
and much larger than in 
the aligned or antialigned cases results in a smaller $m=1$ growth timescale,
as can been seen from the slope of the fitted dashed lines (in the top panel of
Fig. \ref{fig:cm}). These timescales are reported in Table  \ref{tab:mg} second
column and are in broad agreement with other studies
\cite{Korobkin2011,Wessel:2020hvu}. If we denote the growth
of the $m=1$ mode as $e^{t/\GT}$, we find that $\GT/P_c=\{0.5,0.9,0.9,0.7\}$ for
cases A1-A4, confirming that the instability  is indeed dynamical. The two
tilted cases show almost identical growth timescales, even though the disk in
case A3 has almost double the mass of the disk in case A2 while their radial
extent is approximately the same. Note that in
\cite{Kiuchi:2011re,Shibata:2021sau}  it was found that more compact (or more
massive) disks are more subject to the dynamical instability, and when
$M_0/M_{\rm bh}\gtrsim 0.6$ the growth timescale can be smaller than $P_c$. Our
models show that timescales $\lesssim P_c$ are possible with even less massive
disks with $M_0/M_{\rm bh}\sim 0.16$.  This result is not surprising
\cite{Goldreich1986} since our disk models have $\ell=\rm const$ which makes
them more prone to the development of the PPI than the models of
\cite{Kiuchi:2011re,Shibata:2021sau}, which have an nonconstant specific angular
momentum profile.  Given the fact that models A2 and A3 have the same spin
magnitude we conclude that the spin tilt is crucial for the determination of
the growth timescale and can be degenerate with the BH-to-disk mass ratio.

The phase angle of the $m=1$ mode is shown in Fig. \ref{fig:phi1} and the
slopes of the fitted dashed lines (Eq. (\ref{eq:modeome})) provide the
corresponding pattern velocities $\Omega_{p,1}$ that are quoted on Table
\ref{tab:mg}. From this figure one can read easily the time for the
saturation of the PPI. 
In particular for case A1 it is $\approx 5.5 P_c$, for A2 it is  
$\approx 4 P_c$, for A3 it is $\approx 3 P_c$, and for A4 it is  
$\approx 6 P_c$. 
These values are in agreement with the top panel of  Fig. \ref{fig:cm}
and show that the larger the tilt, the smaller the timespan for the development
of the nonaxisymmetric instability.  After this initial period, the mode growth
saturates and the phase angle $\GP_1$ asymptotes to a constant.  Interestingly,
the $m=1$ pattern speed is almost identical for the cases A1 and A2 despite the
different spin orientiations of the BHs, as well as the different BH to disk
mass ratios. This may be related to the fact that those models have identical
inner $r_{\rm in}$ and outer $r_{\rm out}$ boundaries, which play a crucial
role for the explanation of the PPI
\cite{Papaloizou84,Papaloizou85,Zurek1986,Blaes1986}.

Another critical component of the PPI is the corotation
radius $r_{\rm cr}$ through which angular momentum is transferred outwards
\cite{Papaloizou84,Goldreich1986,Zurek1986,Hawley91}. In Table \ref{tab:mg} we
report the ratio of the corotation radius to the radius of the maximum
density for our models A1-A4. This ratio is close to unity, which is typical of
the $m=1$ PPI mode \cite{Korobkin2011,Mewes2015,Mewes2016}. In terms of the
total mass of the system  the corotation radii are $r_{\rm cr}/M=\{16, 17, 17, 17\}$.
In order to confirm and better understand the development of the PPI in thick,
tilted self-gravitating BHDs we plot in Fig. \ref{fig:ell} the specific angular
momentum $\ell=-u_\GP/u_t$ at three different instances for case A2. At one
rotation period (left panel) the disk has essentially the angular momentum
profile  of the initial data i.e. $\ell=\rm const$. After three rotation
periods (middle panel), when the PPI has been well developed, we see two
characteristics: (i) a shock front located at approximately $r \sim 20 M$, and
(ii) the shock front separating the inner part ($r\lesssim 20 M$) of the disk
with angular momentum regions having values smaller than the initial
angular momentum (white-blue areas) from the outer part ($r\gtrsim 20 M$) of
the disk with angular momentum regions having values larger than the
initial angular momentum (green-yellow-red areas). Also, a spiral structure in
the outer part starts to form.  After six rotation periods (right panel), where
the PPI is fully developed, this picture is even clearer and the characteristic
spiral arm is apparent. This shows how the PPI can redistribute angular
momentum by outward transport.

The growth of the one-arm instability results in a pseuso-binary system consisting of
the BH and the $m=1$ ``planet'' that sets the BH in motion. In Fig. \ref{fig:BHpos}
we depict the trajectory of the BH in the equatorial (top panel) and meridional
(bottom panel) planes. In all cases we notice the characteristic spiral
trajectory resulting from the spiral motion of matter in the disk (see
Fig. \ref{fig:rhozoom} left column and Fig. \ref{fig:ell}) and the conservation
of the center of mass of the system.  For case A1 the motion is planar (in the
xy plane) with larger radius of curvature in the beginning when the PPI
develops and smaller at the end, when it has saturated. For the tilted cases A2
and A3 this motion is three dimensional, while for the antialigned case A4 we
again have a three-dimensional motion due to the destabilization of the whole
system after $\sim 6$ rotation periods\footnote{
Note that the linear drift observed in the later part of the A2, A3 orbits
in Fig. \ref{fig:BHpos} may be partly due to the BSSN formalism used in our simulations.
}. The evolution of A4 will be further 
described in the next section.  The combined motion of the BH with the
self-gravitating disk produces copious amounts of gravitational radiation, as we
will discuss next.

\begin{figure}
\begin{center}
\includegraphics[width=0.99\columnwidth]{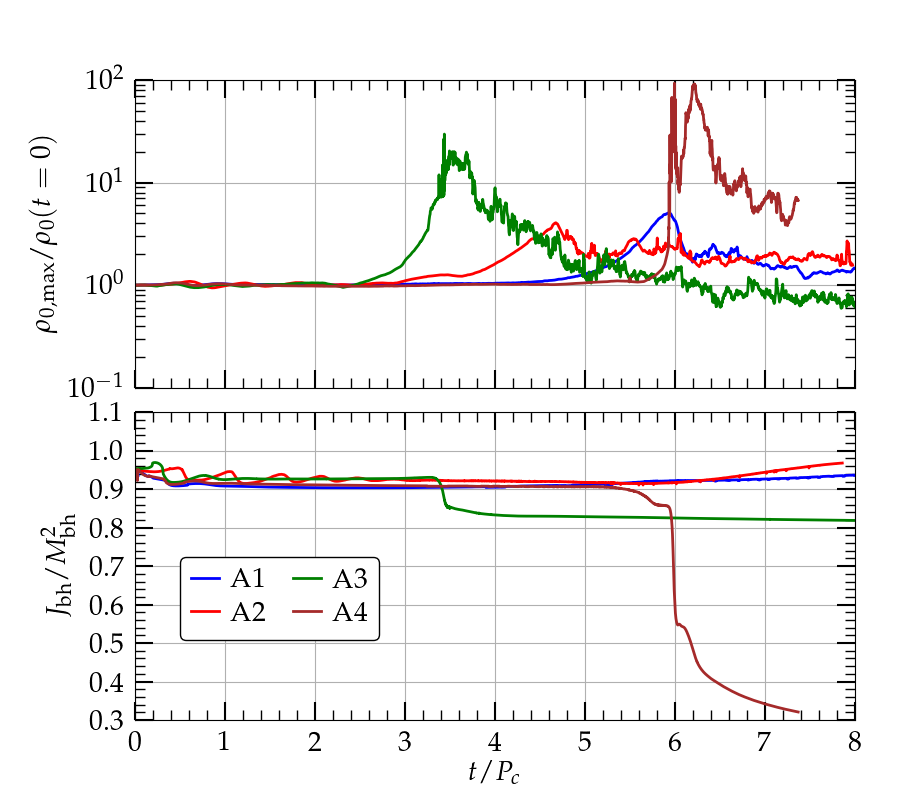}
\caption{Evolution of the maximum rest-mass density of the disk (top panel) and the 
BH dimensionless spin (bottom panel).}
\label{fig:rho_chi}
\end{center}
\end{figure}

\begin{figure}
\begin{center}
\includegraphics[width=0.99\columnwidth]{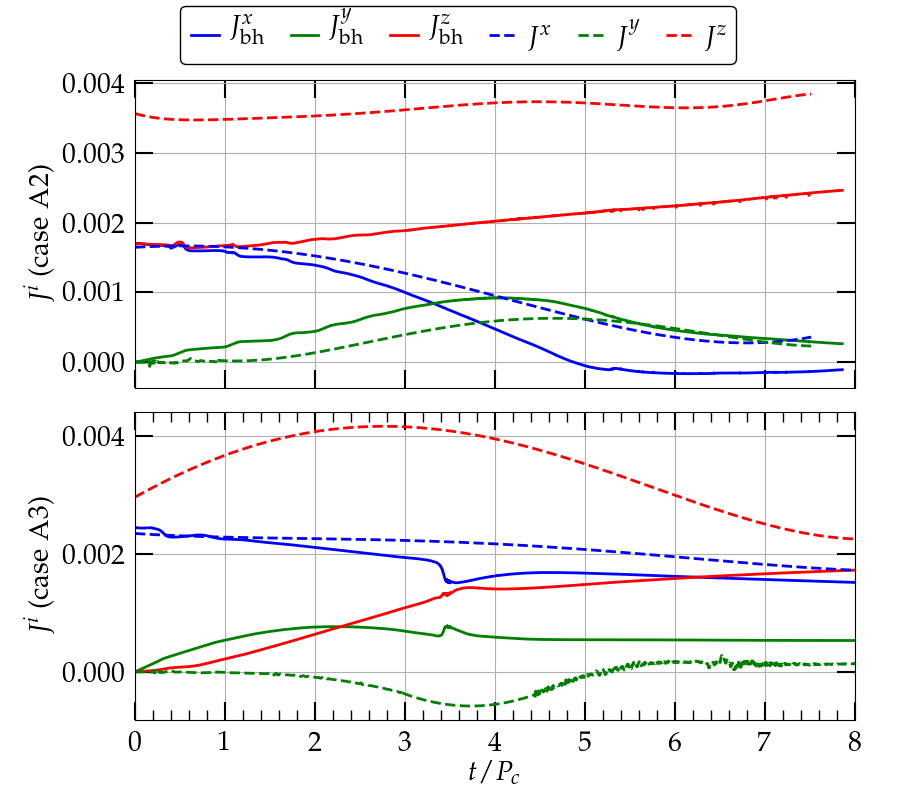}
\caption{Evolution of the BH (solid lines) and ADM (dashed lines) angular momentum components 
for the tilted cases  A2 ($45^\circ$) and A3 ($90^\circ$).}
\label{fig:Jbh_A23}
\end{center}
\end{figure}

\subsection{Precession and gravitational waves}
\label{sec:prec}

In the top panel of Fig. \ref{fig:rho_chi} we plot the evolution of the maximum
density in the disk. The general trend shows the maximum density to be constant
until approximately the end of the development of the PPI, at which point
nonlinear growth sets in and can lead to an increase of $\GR_{0,\rm max}$ by
orders of magnitude.  Consistent with Figs. \ref{fig:cm} and \ref{fig:phi1} we
observe the peak of $\GR_{0,\rm max}$ for case A1 to happen at $\sim 6 P_c$
which coincides with the end of the linear growth of the phase angle $\GP_1$ in
Fig. \ref{fig:phi1}.  Similarly for the cases A2, A3, and A4 the peak times are
$\sim \{4.5 P_c, 3.5 P_c, 6 P_c\}$.  Depending on the characteristics of the
system the maximum density relaxes to values higher or lower than the initial
maximum density and leads to persistent emission of gravitational waves.
Also, as already discussed, the larger the tilt, the earlier the peak of
the maximum density.

\begin{figure}
\begin{center}
\includegraphics[width=0.99\columnwidth]{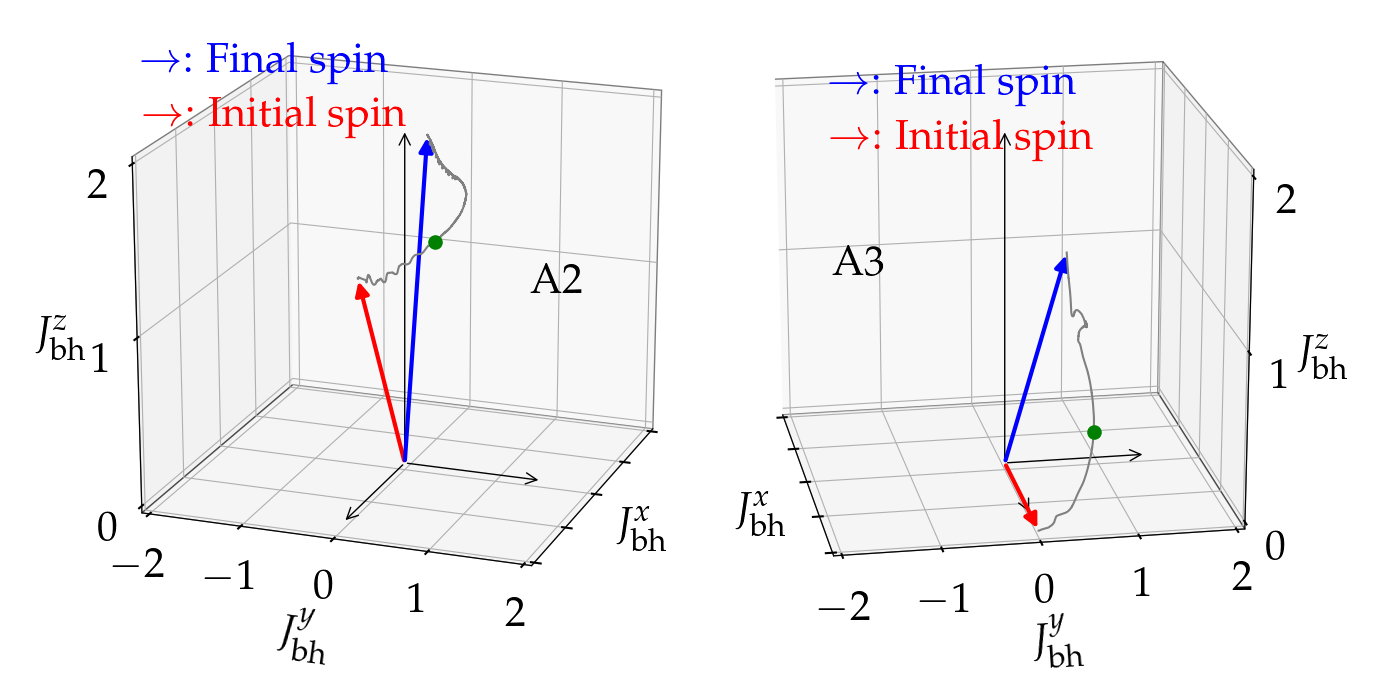}
\caption{BH spin precession for the two tilded cases A2 and A3. The magnitude
of the spin vector is not in scale. The gray curve shows the evolution of the
spin from its initial value (red arrow) to its final value (blue arrow). Green
dots denote times $t=3P_c$ for model A2, and $t=2P_c$ for model A3.}
\label{fig:spinA2A3}
\end{center}
\end{figure}

In the bottom panel of Fig. \ref{fig:rho_chi} the dimensionless spin parameter
$\chi = J_{\rm bh}/M_{\rm bh}^2$ is plotted as a function of time for all our
models.  We adopt the AHFinderDirect thorn
\cite{Thornburg2003:AH-finding_nourl} to locate and monitor the apparent
horizon, and the isolated horizon formalism \cite{dkss03} to measure the mass
of the BH, $M_{\rm bh}$, and its dimensionless spin parameter $\chi$. 
For the cases A1, A4, we have
also confirmed that the Kerr formula for the ratio of proper polar horizon
circumference $L_p$, to the equatorial one $L_e$, $L_p/L_e =
4\sqrt{r_{+}^2+a^2}\ E\left(\frac{a^2}{r_{+}^2+a^2}\right)$ (here $E(x)$ is the
complete elliptic integral of the second kind, $r_{+}$ the event horizon in
Boyer-Linquist coordinates, and $a=J_{\rm bh}/M_{\rm bh}$ the Kerr spin
parameter), and its approximation $L_p/L_e \approx (\sqrt{1-(a/M_{\rm
bh})^2}+1.55)/2.55$ \cite{Brandt1995}, yields almost identical results for the
evolution of $\chi$.  For the tilted case A2 we observe that the BH is spun up
and approaches maximum spin, which prevented us to continue the simulation beyond
$\sim 8$ rotation periods.  For the $90^\circ$ tilted case A3 we observe that
when the maximum density peaks at $\sim 3.5 P_c$ significant accretion onto the
BH is initiated, which results in a reduction of the rest mass of the disk. At
the same time the mass of the BH increases, which leads to an abrupt decrease of
its dimensionless spin to $\chi\sim 0.85$.  By the end of our simulation at
$\sim 8 P_c$ the disk has $75\%$  of its initial mass and the spin of the BH
asymptotes to $\chi\sim 0.82$. The most unstable case in our simulations is
the antialigned case A4. At 6 rotation periods the maximum rest-mass
density increases by two orders of magnitude and shortly afterwards massive
accretion is initiated. That increases the BH mass significantly and its spin drops
to $\sim 0.5$.  Interestingly, the x and y spin components do not show
any appreciable change (i.e. they remain zero), only the z component reduces in
magnitude.  We didn't observe such instability in
\cite{Wessel:2020hvu} where a model with a much smaller spin $\chi=-0.7$ was
employed. We plan to investigate this issue in the future. The
evolution of the three components of the BH spin as well as the
three components of the ADM angular momentum for the two tilted cases A2 and A3
are  plotted in Fig. \ref{fig:Jbh_A23}.  
In Fig. \ref{fig:spinA2A3} we plot the BH spin for the tilted models A2 and A3 
as it evolves from its initial value (red arrow) to its final one (blue arrow). 
The gray curve shows the path of the BH spin vector along our simulations.
In order to verify that precession is observed and measured well before significant
accretion arises, and to measure accurately the GM-induced precession 
we show a green bullet that corresponds to $t=3P_c$ for model A2 and $t=2P_c$ for
model A3. Although at those times the PPI is growing (see Figs.
\ref{fig:cm} and \ref{fig:phi1}) the rest masses of the disks are essentially 
the same as their initial values. The precession of the BH spin from its
initial value (red arrows) to the green bullets is thus mainly due to the 
GM effect. 
Projecting the gray path onto the x-y plane and computing its radius of
curvature we find that the angle between the projections of the initial spin vector and the
spin vector corresponding to the green bullet is $\approx 18^\circ$ or $P_{\rm GM}/20$,
which yields $P_{\rm GM}\approx 60 P_c$. This value exactly matches 
the estimate from the analysis of Section \ref{sec:pf} reported in 
Table \ref{tab:id}. A similar calculation for model A3 yields an angle 
between the projections of the initial spin vector and the
spin vector corresponding to the green bullet of
$\approx 19^\circ$ or $P_{\rm GM}/19$. Hence $P_{\rm GM}\approx 38 P_c$ which
is in excellent agreement with the estimate reported in 
Table \ref{tab:id}. Therefore our simulations are in agreement with the estimates 
from the PN analysis in Section \ref{sec:pf}.

\begin{figure*}
\begin{center}
\includegraphics[width=0.99\columnwidth]{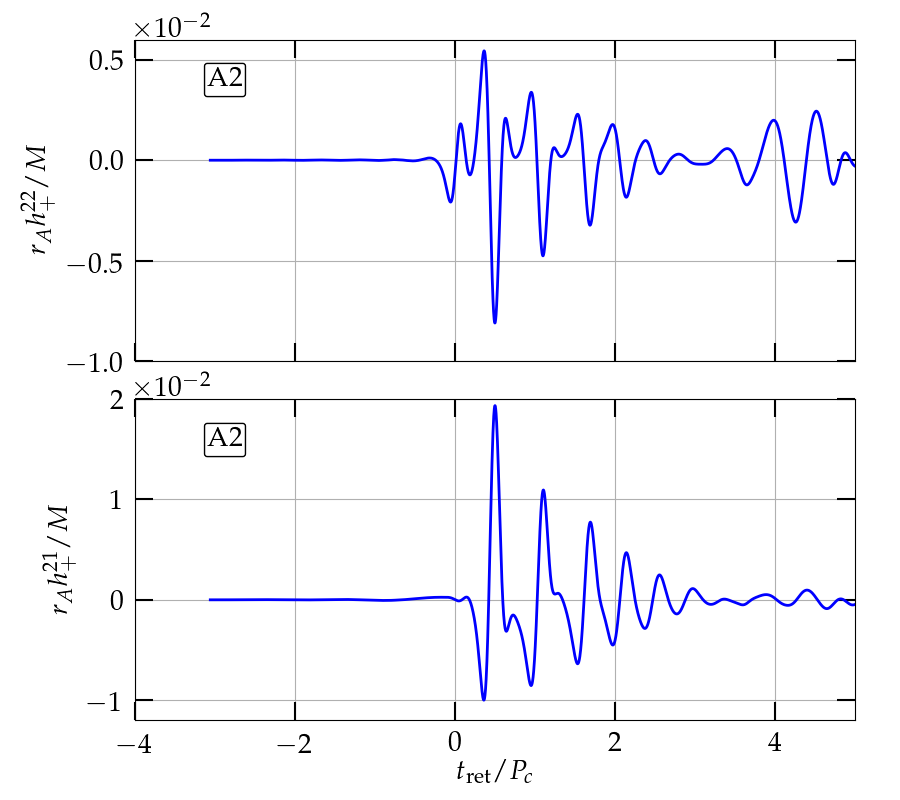}
\includegraphics[width=0.99\columnwidth]{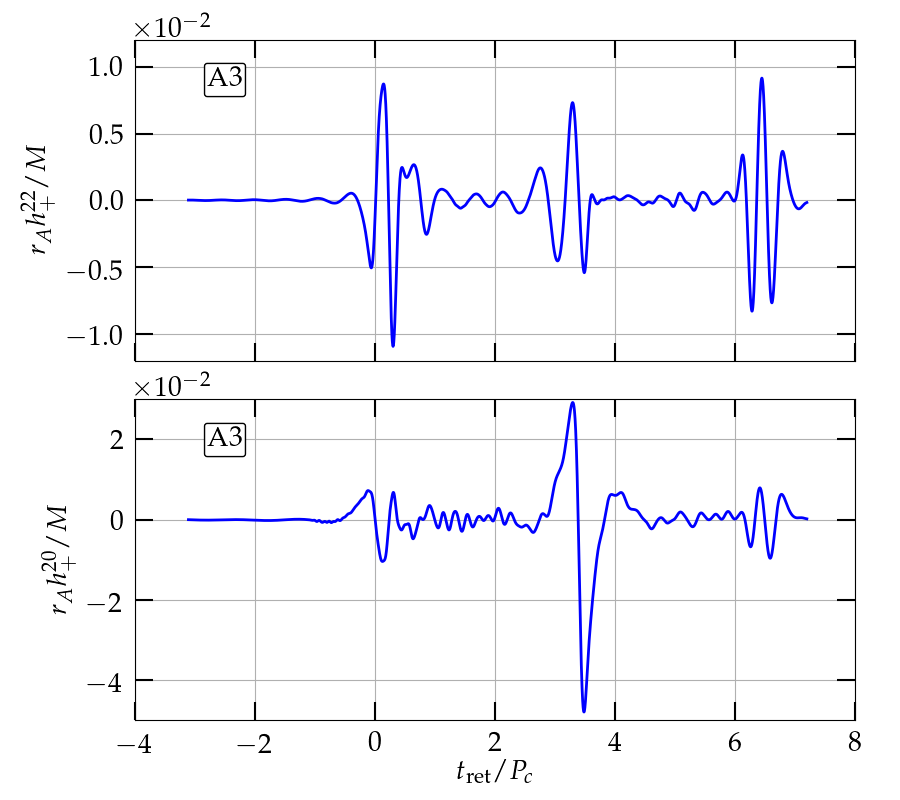}
\caption{Strain amplitude ($h_{+}$) for various gravitational wave modes for the 
two tilted models A2 and A3. Here $r_A$ is the areal extraction radius and 
$t_{\rm ret}$ is retarded time.}
\label{fig:rh}
\end{center}
\end{figure*}

\begin{figure*}
\begin{center}
\includegraphics[width=0.99\columnwidth]{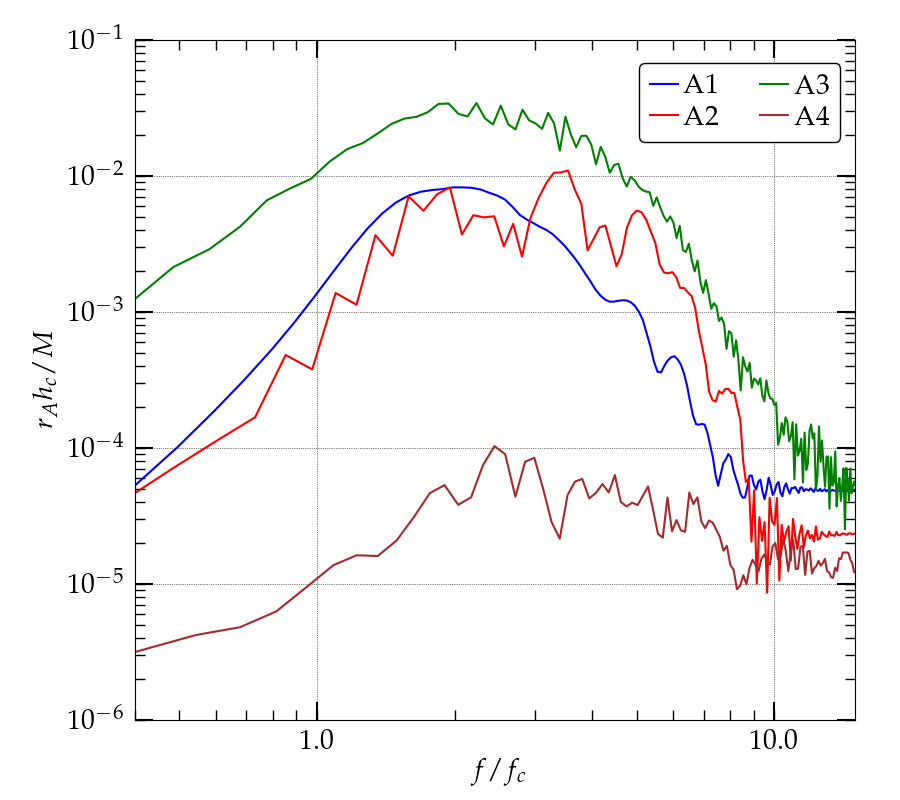}
\includegraphics[width=0.99\columnwidth]{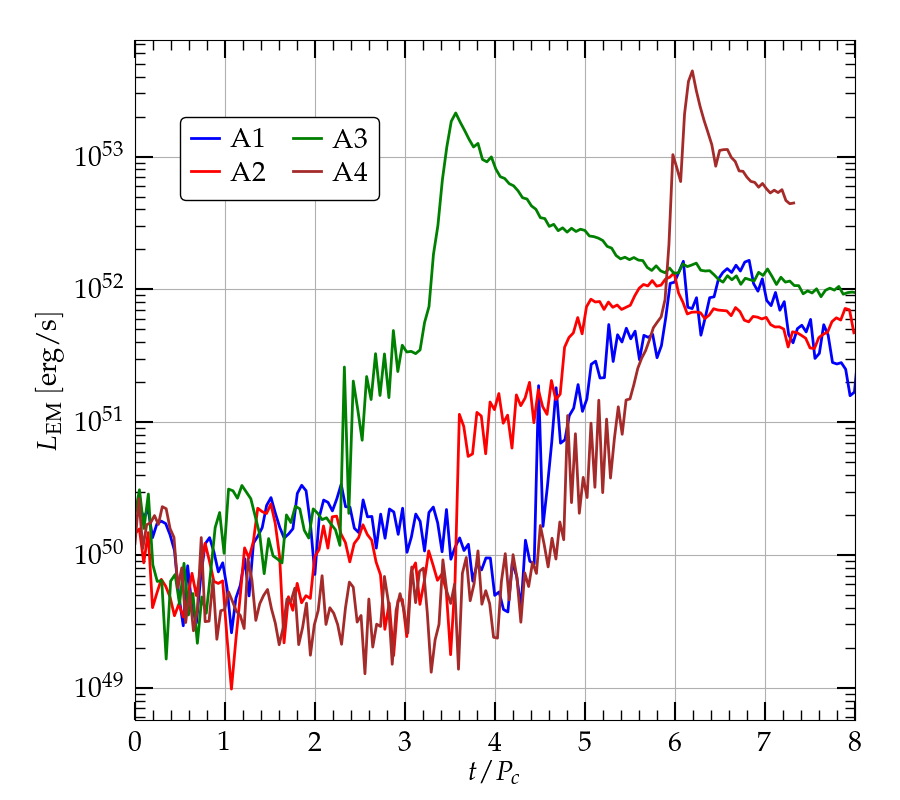}
\caption{Left panel: Gravitational wave spectrum of the $(2,2)$ mode. 
Right panel: Estimated bolometric luminosities. }
\label{fig:spec}
\end{center}
\end{figure*}

\subsection{Multimessenger astronomy}
\label{sec:multi}

BHDs are  prominent sources of electromagnetic radiation due to
accretion.  In our case because of the self-gravity of the disk such systems
also produce significant amounts of gravitational radiation, which makes them
excellent sources for multimessenger astronomy. For the extraction of
gravitational waves we measure the outgoing component of the complex Weyl
scalar $\Psi_4$ expanded in terms of the spin-weighted spherical harmonics with
spin weight $-2$. at various finite radii. The axis of the spherical harmonics
is taken to be the z-axis which is the initial direction of the disk angular
momentum.  The strain $h$ is then computed with a double integration in time as
described in \cite{Reisswig:2011}.

In previous studies
\cite{Kiuchi:2011re,Mewes2015,Wessel:2020hvu,Shibata:2021sau}, where nonspinning
or aligned BHD systems were analyzed, it was found that the development and
saturation of the PPI leads to an initial wave burst, and then a relaxation to
a persistent quasimonochromatic signal of lower amplitude. The peak amplitude
of the strain depends on the disk-to-BH mass ratio as well as the disk
characteristics. Disks of constant specific angular momentum profiles develop a
more pronouced $m=1$ instability, thus the amplitude of gravitational wave is
larger. As explained in \cite{1994ApJ...423..344L,Wessel:2020hvu} it is $rh
\sim O( (r_c \Omega_c)^2)$ and therefore the amplitude of the strain is
directly related to the angular velocity and radius of the maximum density
point. 

When the orbital angular momentum and the BH spin are misaligned this will
cause the precession of the orbit and a modulation of the gravitational waves.
As we have seen in Section \ref{sec:pf}, the angular velocity of the orbital
precession is much smaller than the orbital angular velocity, which implies
that we will need many rotation periods to observe the imprint of precession
on the gravitational waves.
In the left column of Fig. \ref{fig:rh} we plot the $(2,2)$
mode (top panel) and $(2,1)$ mode (bottom panel) of $h_{+}$ for the tilted case
A2 ($r_A$ is the areal extraction radius).  As we discussed above we could not
evolve this model beyond 8 rotation periods due to the almost extremal spin the
BH acquires from accretion. Despite that we observe that the initial
amplitude of the strain is much larger than in the aligned cases (see for
example \cite{Wessel:2020hvu}).  In this particular model the $(2,1)$ mode has
a larger initial amplitude than the $(2,2)$ mode.  The reason for this large
initial amplitude is not due to the $r_c \Omega_c$ value mentioned above but 
from the large nonaxisymmetry of the system at $t=0$. Indeed, the
aligned model A1 has the same $r_c \Omega_c$ value as model A2 but it has a
much smaller peak strain even though the rest mass of the disk is larger. 

Similar large amplitudes are found on the right panels of Fig. \ref{fig:rh}
where the $h_{+}$ strain of the modes $(2,2)$ and $(2,0)$ are plotted for the
tilted case A3. The large peak of the $(2,0)$ mode is also present in the
$(2,1)$ mode, characteristic of mode mixing.  Contrary to the A2 case where the
$\ell=3$ modes are negligible, case A3 has significant amplitude $\ell=3$
modes.  In \cite{Mewes2016} where spins up to $\chi\sim 0.5$ and tilt angles up
to $\sim 30^\circ$ were employed it was found that the gravitational wave
signal has a weak dependence on the initial tilt angle, especially for disks
with nonconstant specific angular momentum profiles. The authors observed the
smallest peak amplitudes for the most tilted BH spacetime. By contrast, in 
our simulations
we see that the gravitational wave signal can be greatly influenced by the
tilt angle as discussed above for the cases A2 and A3. Also, for case A3, which
has the largest tilt we observe the largest peak amplitude. 

We compute the Fourier power spectrum of the gravitational waves 
for the $(2,2)$ mode by calculating
\be
\tilde{h}(f) = \sqrt{\frac{|\tilde{h}^{22}_{+}(f)|^2 +|\tilde{h}^{22}_{\times}(f)|^2}{2}} .
\label{eq:hf}
\ee
Here $\tilde{h}^{22}_{+}(f)$ and  $\tilde{h}^{22}_{\times}(f)$ are the Fourier transforms
of the two independent polarizations $+$ and $\times$. In left panel of Fig.
\ref{fig:spec} we plot the dimensionless characteristic strain $h_c(f) = 2 f
\tilde{h}(f)$ for the four models A1-A4. Case A1 and A3 have peaks at twice the
orbital frequency $f_c$ while A2 at approximately $3f_c$ and a secondary one at
$2f_c$.  The short evolution of the latter, due to reaching maximal spin,
reflects mainly the initial spectral content for that model, i.e. for $t_{\rm
ret}\lesssim 2 P_c$ in left panels of Fig.  \ref{fig:rh}, where a modulation of
the gravitational wave is present. For $t_{\rm ret}\gtrsim 2 P_c$ this
modulation is smoothed out. We expect that this effect is due to the specific
structure of the BHD. As explained in detail in \cite{Wessel:2020hvu} the
gravitational waves depend on the mass of the system from which they
originate and will be excellent sources for the future gravitational wave
observatories. 
In addition, for tilted BHDs the gravitational wave strain of modes beyond the 
$(2,2)$ mode is as strong as the $(2,2)$ one (see Fig. \ref{fig:rh} bottom row), thus 
the magnitude of their characteristic strain will be comparable with that of 
Fig. \ref{fig:spec} (left panel) and therefore detectable by future 
gravitational wave observatories.

In the presence of magnetic fields simulations of compact objects that lead to
the formation of BHDs have shown that they can power relativistic jets
\cite{Paschalidis:2014qra,Ruiz:2016rai,Ruiz:2017due,Ruiz:2018wah,Ruiz:2019ezy,Ruiz2021,Sun2022}
with an outgoing electromagnetic Poynting luminosity of $L_{\rm EM} \sim
10^{52\pm 1}\ \rm erg/s$.  These relativistic jets are consistent with the
Blandford-Znajek mechanism for launching jets and their associated Poynting
luminosities \cite{BZeffect}.  Although our simulations are lacking magnetic
fields we can still estimate the Poynting electromagnetic luminosity, since the power
available for electromagnetic jet emission is usually proportional to the accretion
power \cite{Shapiro83}, i.e.
\be
L_{\rm EM} = \GE \dot{M}_0 c^2 \ ,
\label{eq:Lem}
\ee
where $\dot{M}_0$ the rest-mass accretion rate and $\GE$ an efficiency factor 
$O(10^{-3})$ to $O(10^{-2})$. Assuming $\GE=0.003$ as in \cite{Ruiz:2020via} we 
plot in the right panel of Fig. \ref{fig:spec} the electromagnetic luminosity
coming out from models A1-A4. The
tilted cases A2, A3 exhibit episodes of accretion at earlier times,
due to the tilted geometry of the ISCO. The 
larger the tilt, the earlier these episodes appear ($2.5 P_c$ for A3 while $3.5 P_c$
for A2). Following these periods, accretion continues to grow exponentially 
until approximately the saturation of the PPI, at which point it drops. 
The tilt seems to affect the asymptotic value 
of the accretion rate. Although longer simulations are needed for more
conclusive results, with radiative transport and magnetic fields 
incorporated, our simulations show that case A3 asymptotes to a larger
value than case A2, which in turn asymptotes to a larger value than case A1,
with the differences being less than an order of magnitude.
From Fig. \ref{fig:spec} we compute the accretion timescale of our
models to be $t_{\rm accr} \approx 2\times 10^{4}-10^{5}\ \mbh$ consistent with
\cite{Kiuchi:2011re,Wessel:2020hvu}. Analogous to the accretion rate,
the accretion timescales follow
$t_{\rm accr}(A1) > t_{\rm accr}(A2) > t_{\rm accr}(A3)$.

On the other hand, the inclusion of magnetic fields will lead to the
development of the magnetorotational instability \cite{Balbus1991} as well as
turbulence \cite{Bugli2018}.  The increase of turbulent viscosity will
redistribute the angular momentum in the disk with the possibility of
suppressing the PPI. Despite this, if the turbulent viscous timescale is much longer
than the timescale for the growth and saturation of the PPI there may be
sufficient time for a multimessenger event. We estimate the viscous timescale
as
\be
\frac{\tau_{\rm vis}}{P_c} = \frac{R^2}{P_c \nu} \approx
 \frac{1}{2\pi \GA_{\rm SS}} \frac{\Omega_c R^2}{c_s H} 
\label{eq:tvis}
\ee
where $\nu = \GA_{\rm SS}H c_s$ is the shear viscosity, ($H,\ R$) the (height,
width) of the disk, $c_s$ the sound speed, and $ \GA_{\rm SS}$ the
Shakura–Sunyaev viscosity parameter \cite{Shakura1973}. In our case $c_s^2 = \Gamma
(\Gamma-1)P/((\Gamma-1)\GR + \Gamma P)$.  For $\GA_{\rm SS}=0.01$ it turns out
that our models have $\tau_{\rm vis}/P_c \sim \{198, 198, 201, 176\}$.  Even if
$\GA_{\rm SS}$ is five times larger, the viscous timescale will be $\sim 40
P_c$ i.e. much larger than the time for PPI development and saturation. This is
especially true for the tilted BHDs, in which case the PPI grows much earlier
than in the aligned/antialigned ones. Therefore our preliminary conclusion is
that the one-arm instability in BHD systems can still be a source for
multimessenger astronomy. Full general relativistic magnetohydrodynamic
simulations with radiative transport will be needed to assess reliably the outcome of such systems.

\section{Discussion}

In this work we initiated a study of tilted, self-gravitating disks around
spinning black holes. Our general relativistic, hydrodynamics simulations
are the first that start from self-consistent initial values and include highly
spinning black holes. In these preliminary simulations we focused on BHDs that
have a constant specific angular momentum profile and the disk to BH mass
ratio is $16\%-28\%$.  We investigated aligned ($0^\circ$), antialigned 
($180^\circ$), and highly tilted systems ($45^\circ$ and $90^\circ$),
all of them having dimensionless spins of $0.96-0.97$. 
The nonaxisymmetric mode analysis showed that the saturation of the
PPI happens earlier than in the aligned/antialigned cases and the $m=1$ mode
growth is smaller. The disks precess and warp around the BHs, which also precess
following PN GM precession periods. This causes the BH
center to acquire a small kick velocity. We confirmed
that after outward angular momentum transport is initiated close to the $m=1$
corotation radius, the disk's maximum density increases (sometimes by
orders of magnitude). Accretion on the BH causes its dimensionless spin either
to increase or to decrease, depending on the configuration. Tilted systems
exhibit earlier accretion episodes than the aligned/antialigned ones.
We also observe a weak dependence on the BH tilt, with larger tilts leading to
higher accretion rates, although
longer simulations are needed.  Gravitational waves from tilted BHDs 
typically have larger strains than the ones coming from aligned/antialigned systems
and exhibit a diverse spectrum of modes beyond the (2,2) mode. We expect such
self-gravitating disks to be excellent sources for multimessenger astronomy.

\begin{acknowledgments}                                                                                                                                                        
We thank members of the Illinois Relativity Undergraduate
Research Team (M. Kotak, J. Huang, E. Yu, and J. Zhou) for assistance with some
of the visualizations. 
This research was supported, in part, by a grant from the Office of Undergraduate 
Research at the University of Illinois at Urbana-Champaign.
This work was supported by National Science Foundation Grant PHY-2006066 and the 
National Aeronautics and Space Administration (NASA) Grant 80NSSC17K0070 to the 
University of Illinois at Urbana-Champaign, 
and NSF Grants PHY-1912619 and PHY-2145421 to the University of Arizona.
M.R. acknowledges also support by the Generalitat Valenciana
Grant CIDEGENT/2021/046. 
This work made use of the
Extreme Science and Engineering Discovery Environment (XSEDE), which is
supported by National Science Foundation Grant TG-MCA99S008.  This research is
part of the Frontera computing project at the Texas Advanced Computing Center.
Frontera is made possible by National Science Foundation award OAC-1818253.
Resources supporting this work were  also provided by the NASA High-End
Computing Program through the NASA Advanced Supercomputing Division at Ames
Research Center.
\end{acknowledgments}

\bibliographystyle{apsrev4-1}
\bibliography{references}

\end{document}